# Analytical solutions for a charged particle with white, thermal, and active noises in the presence of a uniform magnetic field

Yun Jeong Kang [a], Sung Kyu Seo [b,] and Kyungsik Kim [c,*]

[a] *School of Liberal Studies, Wonkwnag University, 54538, Republic of Korea*

[b] *Haena Ltd., Seogwipo-si, Jejudo 63568, Republic of Korea*

[c] *Department of Physics, Pukyong National University, Busan 48513, Republic of Korea*

In this paper, we apply the double Fourier transform method to the two-dimensional Vlasov equations for a charged particle subjected to white, thermal, and active noises in uniform a magnetic field. By deriving the corresponding Fokker–Planck equation, analytical solutions for the joint probability density are obtained in different time domains. The mean squared velocity of a charged particle driven by white noise exhibits a superdiffusive behavior, scaling as $\sim t^2$ in the short-time domain, while it grows linearly with time ($\sim t$) in the long-time domain, in agreement with numerical simulations of the mean squared displacement. When thermal noise is included together with trap and viscous forces, the characteristic time scale increases as $\sim t^{2h+1}$ in the corresponding time regimes, whereas the mean squared velocity scales as $\sim t^{2h+3}$. The moments of the joint probability density under thermal noise scale as $\sim t^{2h+5}$. Furthermore, in the limit $h \to 1/2$, the entropy of the joint probability density associated with thermal noise coincides with that obtained for active noise in both the short-time ($t \ll \tau$) and long-time ($t \gg \tau$) limits.

## 1.Introduction

The Vlasov–Fokker–Planck equation is a fundamental partial differential equation obtained by combining the Vlasov equation [1], which describes collisionless particle dynamics under long-range electromagnetic forces, with the Fokker–Planck equation, which models stochastic diffusion processes in velocity space. This equation governs the evolution of particle distribution functions in phase space for systems in which collective self-consistent fields coexist with weak collisions or dissipative effects. The Vlasov equation has been widely applied in statistical mechanics, plasma physics, particle beam dynamics, and even in certain biological systems.

Prior to the last four decades, the Chirikov–Taylor model [2,3] was extensively studied, and analytical solutions were developed for a wide range of plasma physics problems [4–6]. Probabilistic methods and powerful analytical techniques for solving the Vlasov equation were directly applied to this model. As a result, analytical expressions for the probability density function, its second velocity moment, and the associated spatial diffusion coefficient were obtained. In the framework of the Vlasov equation, a collisionless plasma is described by deterministic particle trajectories in a magnetic field. Each particle follows a spiral trajectory around the magnetic field lines determined by its initial conditions. Consequently, the mean squared displacement either oscillates around a constant value or converges to a finite limit set by the initial distribution, rather than diverging indefinitely. This behavior reflects the fact that particles remain confined by the magnetic field and do not undergo unbounded diffusion.

Jimenez-Aquino and Romero-Bastida [7] investigated diffusion in the Langevin discharge equation, incorporating the effects of magnetic fields and colored noise, and analyzed long-time diffusion behavior as well as the influence of magnetic fields on the mean squared displacement. Tothova and Lisy [8] studied the diffusion of charged particles using an exponentially correlated random force model, identifying distinct time-domain behaviors through linear and exponentially decaying contributions to the mean squared displacement. In a subsequent work, Tothova and Lisy [9] showed that subdiffusive behavior ($\alpha = 1/2$) may emerge in the long-time limit within a generalized Langevin equation framework. More recently, we obtained analytical probability densities for active particles coupled to two heat reservoirs [10], passive particles subjected to harmonic and viscous forces [11], and passive particles in random environments under magnetic fields with correlated Gaussian forces [12]. In addition, we analyzed the dynamical behavior of an active Brownian particle confined by an optical trap and driven by correlated Gaussian noise [13].

The modified Vlasov model that includes magnetic fields is of particular interest because it can be extended to chaotic and turbulent stochastic dynamics. This extension becomes possible through the incorporation of correlated Gaussian forces, which allows new analytical solutions to be obtained using the Fourier transform method. The purpose of this paper is to present analytical solutions of the modified Vlasov equation. Specifically, we consider a charged particle in a magnetic field subjected to exponentially correlated Gaussian forces. The paper is organized as follows. In Section 2, we introduce the modified Vlasov equation for a charged particle in a magnetic field and derive the probability density functions for the displacement and velocity in the limits $t \ll \tau$, $t \gg \tau$ and for $\tau = 0$, where $\tau$ denotes the correlation time. In Section 3, we analyze the Vlasov equation with exponentially correlated Gaussian forces in a magnetic field. Section 4 presents analytical solutions for a charged colloidal particle subjected to trap forces, thermal equilibrium noise $\zeta_{th}(t)$, and active noise $\zeta_{ac}(t)$ with exponentially decaying correlations. In Section 5, we compute statistical quantities such as the non-Gaussian parameter, correlation coefficients, entropy, and combined entropy. Finally, our conclusions are summarized in Section 6.

## 2. Modified Vlasov equation

The general Vlasov equation [6] including nonrelativistic electric and magnetic fields is given by

$$\frac{\partial f}{\partial t} + \boldsymbol{v} \cdot \nabla_r f + \frac{q}{m}(\boldsymbol{E} + \boldsymbol{v} \times \boldsymbol{B}) \cdot \nabla_v f = D \nabla_r^2 f. \tag{1}$$

In the case where only the magnetic field $\boldsymbol{B} = B_z \hat{z}$ exists in two dimensions, the probability density function in the Vlasov equation is written as

$$\frac{\partial}{\partial t} f + v_x \frac{\partial}{\partial x} f + v_y \frac{\partial}{\partial y} f + \frac{q}{m} v_y B_z \frac{\partial}{\partial v_x} f - \frac{q}{m} v_x B_z \frac{\partial}{\partial v_y} f = D_x \frac{\partial^2}{\partial x^2} f + D_y \frac{\partial^2}{\partial y^2} f. \tag{2}$$

Here, the velocity is $\boldsymbol{v} = v_x \hat{x} + v_y \hat{y}$, and $D_x$ and $D_y$ denote the spatial diffusion coefficients. The equations of motion for a charged particle in the presence of $\boldsymbol{B} = B_z \hat{z}$ are given by

$$\frac{\partial}{\partial t} x = v_x + \eta_x(t), \tag{3}$$

$$m \frac{\partial}{\partial t} v_x = -q v_y B_z, \tag{4}$$

and

$$\frac{\partial}{\partial t} y = v_y + \eta_y(t), \tag{5}$$

$$m \frac{\partial}{\partial t} v_y = +q v_x B_z. \tag{6}$$

Here, the random forces $\eta_x(t)$ and $\eta_y(t)$ depend on the time difference and satisfy $<\eta_i(t)\eta_i(t')> = \frac{D_i}{\tau}\delta(\frac{|t-t'|}{\tau})$ for $i = x, y$. We show in Appendix A that the time derivative of the joint probability density derived from Eqs. (3)-(6) is consistent with Eq. (2).

The joint probability densities $f(x, v_x, t)$ and $f(y, v_y, t)$ for the displacement $x, y$ and the velocity $v_x, v_y$ are defined by $f(x, v_x, t) = <\delta(x - x(t))\delta(v_x - v_x(t))>$ and $f(y, v_y, t) = <\delta(y - y(t))\delta(v_y - v_y(t))>$. By taking time derivatives of these joint probability densities, we have the differential equations

$$\frac{\partial}{\partial t} f(x, v_x, t) = -\frac{\partial}{\partial x} <\frac{\partial x}{\partial t}\delta(x - x(t))\delta(v_x - v_x(t))> - \frac{\partial}{\partial v_x} <\frac{\partial v_x}{\partial t}\delta(x - x(t))\delta(v_x - v_x(t))>, \tag{7}$$

$$\frac{\partial}{\partial t} f(y, v_y, t) = -\frac{\partial}{\partial y} <\frac{\partial y}{\partial t}\delta(y - y(t))\delta(v_y - v_y(t))> - \frac{\partial}{\partial v_y} <\frac{\partial v_y}{\partial t}\delta(y - y(t))\delta(v_y - v_y(t))>. \tag{8}$$

we then introduce the following equations for the evolution of the joint probability densities:

$$\frac{\partial}{\partial t} f(x, v_x, t) = [-v_x \frac{\partial}{\partial x} + v_x B_z^2 \frac{\partial}{\partial v_x}] f(x, v_x, t) + D_x \frac{\partial^2}{\partial x^2} f(x, v_x, t) \tag{9}$$

$$\frac{\partial}{\partial t} f(y, v_y, t) = [-v_y \frac{\partial}{\partial y} + v_y B_z^2 \frac{\partial}{\partial v_y}] f(y, v_y, t) + D_y \frac{\partial^2}{\partial y^2} f(y, v_y, t). \tag{10}$$

Here, the dimensionless parameters $m = 1$ and $q = 1$ are used. The derivation of Eqs. (9) and (10) is presented in Appendix B.

$$f(\zeta_1, v_1, t) = \int_{-\infty}^{+\infty} dx \int_{-\infty}^{+\infty} dv_x \, exp(-i\zeta_1 x - iv_1 v_x) f(x, v_x, t), \qquad (11)$$

$$f(\zeta_2, v_2, t) = \int_{-\infty}^{+\infty} dy \int_{-\infty}^{+\infty} dv_y \, exp(-i\zeta_2 y - iv_2 v_y) f(y, v_y, t). \qquad (12)$$

By applying the double Fourier transform to Eqs. (9) and (10), we obtain the variable-separating equations

$$\frac{\partial}{\partial t} f(\zeta_1, v_1, t) = [\zeta_1 - B_z^{\,2} v_1] \frac{\partial}{\partial v_1} f(\zeta_1, v_1, t) - D_x v_1^2 f(\zeta_1, v_1, t) + A f(\zeta_1, v_1, t), \qquad (13)$$

$$\frac{\partial}{\partial t} f(\zeta_2, v_2, t) = [\zeta_2 - B_z^{\,2} v_2] \frac{\partial}{\partial v_2} f(\zeta_2, v_2, t) - D_y v_1^2 f(\zeta_2, v_2, t) - A f(\zeta_2, v_2, t), \qquad (14)$$

where $A$ is the separation constant, and the initial conditions are $x_0 = y_0 = 0$ and $v_{x0} = v_{y0} = 0$.

*2.1. 2D Probability Densities in steady state*

In this subsection, we solve the Fourier transforms of the probability density function given by Eq. (13) and Eq. (14) for the modified Vlasov equation, Eq. (9), describing a charged particle.

2.1.1. $f(x,t)$ and $f(v_x,t)$ in the short- and long-time domains

By taking $\frac{\partial}{\partial t} f(\zeta_1, v_1, t) = 0$ for $v_1$ in the steady state, $f(\zeta_1, v_1, t)$ reduces to the stationary distribution $f^{st}(\zeta_1, v_1, t)$. The Fourier transformed equation for the steady probability density from Eq. (13) is

$$[\zeta_1 - B_z^{\,2} v_1] \frac{\partial}{\partial v_1} f^{st}(\zeta_1, v_1, t) - D_x \zeta_1^2 f^{st}(\zeta_1, v_1, t) + A f^{st}(\zeta_1, v_1, t) = 0. \qquad (15)$$

The first term of the left-hand side in Eq. (15) is approximated as $[\zeta_1 - B_z^{\,2} v_1]^{-1} = \zeta_1^{-1}[1 + \frac{B_z^{\,2} v_1}{\zeta_1}]$. The stationary solution with respect to $v_1$ is then given by

$$f^{st}(\zeta_1, v_1, t) = exp[\frac{1}{\zeta_1}[D_x \zeta_1^2 v_1 - A v_1] + \frac{B_z^{\,2}}{\zeta_1^{\,2}}[D_x \zeta_1^2 \frac{v_1^2}{2} - A v_1]]. \qquad (16)$$

Introducing $f(\zeta_1, v_1, t) = \Theta[t + v_1/[\zeta_1 - B_z v_2]] f^{st}(\zeta_1, v_1, t)$, we obtain

$$f(\zeta_1, v_1, t) = exp[-D_x t \zeta_1^2 - D_x B_z^{\,2} t^2 \zeta_1^2/2 - D_x B_z^{\,6} t^2 v_1^2/2]. \qquad (17)$$

Using the inverse Fourier transform, the probability densities, in the short-time domain when $D_x t \ll 1$ are given by

$$f(x,t) = [2\pi D_x B_z^{\,2} t^2]^{-1/2} exp[-\frac{x^2}{2 D_x B_z^{\,2} t^2}], f(v_x, t) = [2\pi D_x B_z^{\,6} t^2]^{-1/2} exp[-\frac{v_x^2}{2 D_x B_z^{\,6} t^2}]. \qquad (18)$$

In the long-time domain when $D_x B_z^{\,2} t^2 \ll 1$, the probability densities are obtained as

$$f(x,t) = [4\pi D_x t]^{-1/2} exp[-\frac{x^2}{4 D_x t}], f(v_x, t) = [2\pi D_x B_z^{\,6} t^2]^{-1/2} exp[-\frac{v_x^2}{2 D_x B_z^{\,6} t^2}]. \qquad (19)$$

Thus, the mean squared quantities corresponding to Eqs. (18) and (19) are given by

$$< x^2(t) > = D_x B_z^{\,2} t^2, \quad < v_x^2(t) > = D_x B_z^{\,6} t^2, \qquad (20)$$

in the short-time domain, and

$$< x^2(t) > = 2 D_x t, \quad < v_x^2(t) > = D_x B_z^{\,6} t^2, \qquad (21)$$

in the long-time domain.

2.1.2. $f(y,t)$ and $f(v_y,t)$ in the short- and long-time domains

Taking $\frac{\partial}{\partial t} f(\zeta_2, v_2, t) = 0$ for $v_1$ in the steady state, $f(\zeta_2, v_2, t)$ becomes $f^{st}(\zeta_2, v_2, t)$. The Fourier transformed equation for the steady probability density function from Eq. (10) is obtained as

$$[\zeta_2 - B_z^2 v_2]\frac{\partial}{\partial v_1} f^{st}((\zeta_2, v_2, t) - D_x \zeta_1^2 f^{st}((\zeta_2, v_2, t) - A f^{st}((\zeta_2, v_2, t) = 0. \tag{22}$$

For Eq. (22), by a similar procedure of Eqs. (16) and (17), we have

$$f(\zeta_2, v_2, t) = exp[-D_y t \zeta_2^2 - D_y B_z^2 t^2 \zeta_2^2/2 - D_y B_z^6 t^2 v_2^2/2]. \tag{23}$$

Using the inverse Fourier transform, the probability density in the short-time domain when $D_y t \ll 1$ is calculated as

$$f(y,t) = [2\pi D_y B_z^2 t^2]^{-1/2} exp[-\frac{y^2}{2D_y B_z^2 t^2}], f(v_y, t) = [2\pi D_y B_z^6 t^2]^{-1/2} exp[-\frac{v_y^2}{2D_y B_z^6 t^2}]. \tag{23}$$

In the long-time domain when $D_y B_z^2 t^2 \ll 1$, the probability density also is calculated as

$$f(y,t) = [4\pi D_y t]^{-1/2} exp[-\frac{y^2}{4D_y t}], f(v_y, t) = [2\pi D_y B_z^6 t^2]^{-1/2} exp[-\frac{v_y^2}{2D_y B_z^6 t^2}]. \tag{24}$$

Thus, the mean squared quantities for $f(y,t)$ and $f(v_y,t)$ from Equations (16) and (17) are, respectively, given by

$$<y^2(t)> = D_y B_z^2 t^2, \ <v_y^2(t)> = D_y B_z^6 t^2, \tag{25}$$

in the short-time domain, and

$$<y^2(t)> = 2D_y t, \ <v_y^2(t)> = D_y B_z^6 t^2, \tag{26}$$

in the long-time domain. Thus, if the diffusion coefficient $D_x$ is equal to $D_y$, then the mean square displacement and mean square velocity in the x- and y-axis directions are equal, as we have seen in Sections 2-1-1 and 2-1-2. This means that the motion of the charged particles has a statistical isotropy in the $x-$ and $y-$axis directions.

## 3. Modified Vlasov equation with exponential correlated Gaussian forces in the magnetic field

In this subsection, we solve the probability density function for the modified Vlasov equation, Eqs. (3)-(6), of a charged particle with two correlated Gaussian forces within the limits of $t \ll \tau$, $t \gg \tau$ and for $\tau = 0$, where $\tau$ is the correlation time.

The two modified equations of motion for a charged particle with $\vec{B} = B_z \vec{z}$ are given by

$$\frac{\partial}{\partial t} x = v_x + g_x(t), \tag{27}$$

$$m\frac{\partial}{\partial t} v_x = -q v_y B_z - r_1 v_x, \tag{28}$$

and

$$\frac{\partial}{\partial t} y = v_y + g_y(t), \tag{29}$$

$$m\frac{\partial}{\partial t} v_y = +q v_x B_z - r_2 v_y. \tag{30}$$

Here, the random forces $g_x(t)$ and $g_y(t)$ depend on the time difference: $<g_i(t) g_i(t')> = \frac{D_i}{\tau} exp(-\frac{|t-t'|}{\tau})$ for $i = x, y$ [13,14]. $-r_1 v_x$ and $-r_2 v_y$ are the viscous forces, respectively.

The joint probability densities $f(x, v_x, t)$ and $f(y, v_y, t)$ for the displacement $x, y$ and the velocity $v_x, v_y$ are defined as $f(x, v_x, t) = <\delta(x - x(t))\delta(v_x - v_x(t))>$ and $f(y, v_y, t) = <\delta(y - y(t))\delta(v_y - v_y(t))>$. By taking time derivatives of these joint probability densities, we obtain the following differential equations:

$$\frac{\partial}{\partial t} f(x, v_x, t) = -\frac{\partial}{\partial x} <\frac{\partial x}{\partial t} \delta(x - x(t))\delta(v_x - v_x(t))> - \frac{\partial}{\partial v_x} <\frac{\partial v_x}{\partial t} \delta(x - x(t))\delta(v_x - v_x(t))>, \tag{31}$$

$$\frac{\partial}{\partial t} f(y, v_y, t) = -\frac{\partial}{\partial y} <\frac{\partial y}{\partial t} \delta(y - y(t))\delta(v_y - v_y(t))> - \frac{\partial}{\partial v_y} <\frac{\partial v_y}{\partial t} \delta(y - y(t))\delta(v_y - v_y(t))>. \tag{32}$$

Substituting Eqs. (27) and (28) into Eq. (31), and Eqs. (29) and (30) into Eq. (32), and following the procedure of

Ref. [13], we obtain

$$\frac{\partial}{\partial t}f(x,v_x,t) = \left[-v_x\frac{\partial}{\partial x} + [-B_z v_{y0}e^{-r_2 t} + (B_z^2+r_1)v_x]\frac{\partial}{\partial v_x}\right]f(x,v_x,t) + D_x[-b_1(t)\frac{\partial^2}{\partial x^2} + a_1(t)\frac{\partial^2}{\partial v_x^2}]f(x,v_x,t), \quad (33)$$

$$\frac{\partial}{\partial t}f(y,v_y,t) = \left[-v_y\frac{\partial}{\partial y} + [B_z v_{x0}e^{-r_1 t} + (B_z^2+r_2)v_y]\frac{\partial}{\partial v_y}\right]f(y,v_y,t) + D_y[-b_2(t)\frac{\partial^2}{\partial y^2} + a_2(t)\frac{\partial^2}{\partial v_y^2}]f(y,v_y,t). \quad (34)$$

Here we have set the dimensionless parameters $m=1$ and $q=1$. The time-dependent coefficients are given by $a_1(t) = a_2(t) = 1 - exp(-t/\tau)$, $b_1(t) = b_2(t) = (t+\tau)exp(-t/\tau) - \tau$. Applying the double Fourier transform defined in Eq. (10), we obtain the Fourier-space equations

$$\frac{\partial}{\partial t}f(\zeta_1,v_1,t) = [\zeta_1 - (B_z^2 + r_1)v_1]\frac{\partial}{\partial v_1}f(\zeta_1,v_1,t) + D_x[b_1(t)\zeta_1 v_1 - a_1(t)\zeta_1^2]f(\zeta_1,v_1,t), \quad (35)$$

$$\frac{\partial}{\partial t}f(\zeta_2,v_2,t) = [\zeta_2 - (B_z^2+r_2)v_2]\frac{\partial}{\partial v_2}f(\zeta_2,v_2,t) + D_y[b_2(t)\zeta_2 v_2 - a_2(t)\zeta_2^2]f(\zeta_2,v_2,t). \quad (36)$$

*3.1. Probability Densities in Three-Time Domains*

3.1.1. $f(x,t)$, $f(v_x,t)$ and $f(y,t)$, $f(v_y,t)$ in the short-time limit $t \ll \tau$

In the short-time domain $t \ll \tau$, the steady-state solution of Eq. (35) for fixed $\zeta_1$ is obtained as

$$f^{st}(\zeta_1,v_1,t) = exp[\int \frac{D_x}{(\zeta_1 - (B_z^2+r_1)v_1)}[b_1(t)\zeta_1 v_1 - a_1(t)\zeta_1^2]dv_1]$$

$$= exp[\frac{D_x}{\zeta_1}[b_1(t)\zeta_1\frac{v_1^2}{2} - a_1(t)\zeta_1^2 v_1] + \frac{D_x(B_z^2+r_1)}{\zeta_1^2}][b_1(t)\zeta_1\frac{v_1^3}{3} - a_1(t)\zeta_1^2\frac{v_1^2}{2}]], \quad (37)$$

which yields the explicit exponential form given in Eq. (37). To proceed, we introduce the successive transformations

$$f(\zeta_1,v_1,t) = g(\zeta_1,v_1,t)f^{st}(\zeta_1,v_1,t), \quad (38)$$

$$g(\zeta_1,v_1,t) = h(\zeta_1,v_1,t)exp[\frac{D_x}{\zeta_1^2}[b_1'(t)\zeta_1\frac{v_1^3}{6} - a_1'(t)\zeta_1^2\frac{v_1^2}{2}] + \frac{D_x(B_z^2+r_1)}{\zeta_1^3}][b_1'(t)\zeta_1\frac{v_1^4}{12} - a_1'(t)(t)\zeta_1^2\frac{v_1^3}{6}]], \quad (39)$$

$$h(\zeta_1,v_1,t) = p(\zeta_1,v_1,t)exp[\frac{D_x}{\zeta_1^3}[b_1''(t)\zeta_1\frac{v_1^4}{12} - a_1''(t)\zeta_1^2\frac{v_1^3}{6}] + \frac{D_x(B_z^2+r_1)}{\zeta_1^4}][b_1''(t)\zeta_1\frac{v_1^5}{60} - a_1''(t)(t)\zeta_1^2\frac{v_1^4}{12}]], \quad (40)$$

$$p(\zeta_1,v_1,t) = q(\zeta_1,v_1,t)exp[\frac{D_x}{\zeta_1^4}[b_1'''(t)\zeta_1\frac{v_1^5}{60}] + \frac{D_x(B_z^2+r_1)}{\zeta_1^5}][b_1'''(t)\zeta_1\frac{v_1^6}{360}]]. \quad (41)$$

where higher-order terms proportional to $1/\tau^3$ are neglected. Under this approximation, $q(\zeta_1,v_1,t)$ satisfies

$$\frac{\partial}{\partial t}q(\zeta_1,v_1,t) = [\zeta_1 - (B_z^2+r_1)v_1]\frac{\partial}{\partial v_1}q(\zeta_1,v_1,t), \quad (42)$$

Expanding up to second order in $t/\tau$ and collecting all contributions, we obtain

$$f(\zeta_1,v_1,t) = exp[-\frac{D_x t^3}{2}\zeta_1^2 - \frac{D_x B_z^4 t^3}{2}v_1^2]. \quad (43)$$

The probability densities for displacement and velocity are then given by

$$f(x,t) = \frac{1}{2\pi}\int_{-\infty}^{+\infty}d\zeta_1 exp(+i\zeta_1 x)f(\zeta_1,t) = [2\pi D_x t^3]^{-\frac{1}{2}}exp[-\frac{x^2}{2D_x t^3}], \quad (44)$$

$$f(v_x,t) = \frac{1}{2\pi}\int_{-\infty}^{+\infty}dv_x exp(+iv_1 v_x)f(v_1,t) = [2\pi D_x B_z^4 t^3]^{-1/2}exp[-\frac{v_x^2}{D_x B_z^4 t^3}], \quad (45)$$

with mean squared values

$$<x^2(t)> = D_x t^3, <v_x^2(t)> = D_x B_z^4 t^3. \quad (46)$$

Applying the same procedure to Eq. (36), we obtain

$$f(\zeta_2,v_2,t) = exp[-\frac{D_y t^3}{2}\zeta_2^2 - \frac{D_y B_z^4 t^3}{2}v_2^2], \quad (47)$$

leading to

$$f(y,t) = [2\pi D_y t^3]^{-\frac{1}{2}}exp[-\frac{y^2}{2D_y t^3}], f(v_y,t) = [2\pi D_y B_z^4 t^3]^{-\frac{1}{2}}exp[-\frac{v_y^2}{D_y B_z^4 t^3}]. \quad (48)$$

Consequently,

$$<y^2(t)> = D_y t^3, <v_y^2(t)> = D_y B_z^4 t^3. \tag{49}$$

3.1.2. $f(x,t)$, $f(v_x,t)$ and $f(y,t)$, $f(v_y,t)$ in $t \gg \tau$

In this subsection, we analyze the probability densities $f(\zeta_1, v_1, t)$ and $f(\zeta_2, v_2, t)$ in the long-time domain $t \gg \tau$. In this limit, Eqs. (35) and (36) can be approximated as

$$\frac{\partial}{\partial t} f(\zeta_1, v_1, t) \cong D_x [b_1(t)\zeta_1 v_1 - a_1(t)\zeta_1^2] f(\zeta_1, v_1, t), \tag{50}$$

$$\frac{\partial}{\partial t} f(\zeta_2, v_2, t) \cong D_y [b_2(t)\zeta_2 v_2 - a_2(t)\zeta_1^2] f(\zeta_2, v_2, t), \tag{51}$$

From these equations, we formally obtain

$$f_{\zeta_1}(\zeta_1, v_1, t) = exp[D_x \int [b_1(t)\zeta_1 v_1 - a_1(t)\zeta_1^2] dt], \tag{52}$$

$$f_{\zeta_2}(\zeta_2, v_2, t) = exp[D_y \int [b_2(t)\zeta_2 v_2 - a_2(t)\zeta_1^2] dt]. \tag{53}$$

Using the relations $f(\zeta_1, v_1, t) = g_{\zeta_1}(\zeta_1, v_1, t) f_{\zeta_1}(\zeta_1, v_1, t)$ and $f(\zeta_2, v_2, t) = g_{\zeta_2}(\zeta_2, v_2, t) f_{\zeta_2}(\zeta_2, v_2, t)$, the steady-state factors are obtained as

$$g_{\zeta_1}^{st}(\zeta_1, v_1, t) = exp[-D_x \int [b_1(t)\zeta_1 v_1 - a_1(t)\zeta_1^2] dt], \tag{54}$$

$$g_{\zeta_2}^{st}(\zeta_2, v_2, t) = exp[-D_y \int [b_2(t)\zeta_2 v_2 - a_2(t)\zeta_1^2] dt]. \tag{55}$$

The steady-state solutions $f^{st}(\zeta_1, v_1, t)$ and $f^{st}(\zeta_2, v_2, t)$, given in Eq. (37), are rewritten as

$$f^{st}(\zeta_1, v_1, t) = exp[\frac{D_x}{\zeta_1}[b_1(t)\zeta_1 \frac{v_1^2}{2} - a_1(t)\zeta_1^2 v_1] + \frac{D_x(B_z^2 + r_1)}{\zeta_1^2}][b_1(t)\zeta_1 \frac{v_1^3}{3} - a_1(t)\zeta_1^2 \frac{v_1^2}{2}]], \tag{56}$$

$$f^{st}(\zeta_2, v_2, t) = exp[\frac{D_y}{\zeta_2}[b_2(t)\zeta_2 \frac{v_2^2}{2} - a_2(t)\zeta_2^2 v_2] + \frac{D_y(B_z^2 + r_2)}{\zeta_2^2}][b_2(t)\zeta_2 \frac{v_2^3}{3} - a_2(t)\zeta_2^2 \frac{v_2^2}{2}]]. \tag{57}$$

The Fourier-space solutions can therefore be written as

$$g(\zeta_1, v_1, t) = h(\zeta_1, v_1, t) exp[-D_x \int [b_1(t)\zeta_1 v_1 - a_1(t)\zeta_1^2] dt], \tag{58}$$

$$g(\zeta_2, v_2, t) = h(\zeta_2, v_2, t) exp[-D_y \int [b_2(t)\zeta_2 v_2 - a_2(t)\zeta_2^2] dt]. \tag{59}$$

In the long-time limit, the coefficients satisfy $a_i(t) = 1$, $b_i(t) = -\tau$ and $\int a_i(t) dt = t - \tau$, $\int b_i(t) dt = -\tau t$ for $i = 1,2$. Accordingly, the arbitrary functions become

$$f(\zeta_1, v_1, t) = \Theta[t + v_1/[(\zeta_1 - D_x(B_z^2 + r_1)v_1)]] g_{\zeta_1}^{st}(\zeta_1, v_1, t) f^{st}(\zeta_1, v_1, t), \tag{60}$$

$$f(\zeta_2, v_2, t) = \Theta[t + v_2/[(\zeta_2 - D_x(B_z^2 + r_2)v_2)]] g_{\zeta_2}^{st}(\zeta_2, v_2, t) f^{st}(\zeta_2, v_2, t). \tag{61}$$

After expanding in powers of $t/\tau$ and collecting all terms, we obtain

$$f(\zeta_1, v_1, t) = exp[-\frac{D_x B_z^2 t^4}{3} \zeta_1^2 - \frac{D_x B_z^4 t^3}{2} v_1^2], \tag{62}$$

$$f(\zeta_2, v_2, t) = exp[-\frac{D_y B_z^2 t^4}{3} \zeta_2^2 - \frac{D_y B_z^4 t^3}{2} v_2^2]. \tag{63}$$

Taking the inverse Fourier transforms, we finally obtain

$$f(x,t) = \left[4\pi D_x B_z^2 t^4/3\right]^{-\frac{1}{2}} exp[-\frac{3x^2}{4D_x B_z^2 t^4}], \quad f(v_x, t) = \left[2\pi D_x B_z^4 t^3\right]^{-1/2} exp[-\frac{v_x^2}{2D_x B_z^4 t^3}], \tag{64}$$

$$f(y,t) = \left[4\pi D_y B_z^2 t^4/3\right]^{-\frac{1}{2}} exp[-\frac{3y^2}{4D_y B_z^2 t^4}], \quad f(v_y, t) = [2\pi D_y B_z^4 t^3]^{-1/2} exp[-\frac{v_y^2}{2D_y B_z^4 t^3}]. \tag{65}$$

The mean-square displacement and velocity are therefore given by

$$<x^2(t)> = 2D_x B_z^2 t^4/3, \quad <v_x^2(t)> = D_x B_z^4 t^3, \tag{66}$$

$$<y^2(t)> = 2D_y B_z^2 t^4/3, \quad <v_y^2(t)> = D_y B_z^4 t^3. \tag{67}$$

3.1.3. $f(x,t)$, $f(v_x,t)$ and $f(y,t)$, $f(v_y,t)$ in the white-noise limit $\tau = 0$

For $\tau = 0$, the coefficients reduce to $a_1(t) = a_2(t) = 1$ and $b_1(t) = b_2(t) = 0$. Equations (35) and (36) then become

$$\frac{\partial}{\partial t}f(\zeta_1, v_1, t) = [\zeta_1 - D_x(B_z^2 + r_1)v_1]\frac{\partial}{\partial v_1}f(\zeta_1, v_1, t) - D_x\zeta_1^2 f(\zeta_1, v_1, t), \tag{68}$$

$$\frac{\partial}{\partial t}f(\zeta_2, v_2, t) = [\zeta_2 - D_x(B_z^2 + r_2)v_2]\frac{\partial}{\partial v_2}f(\zeta_2, v_2, t) - D_y\zeta_2^2 f(\zeta_2, v_2, t). \tag{69}$$

The steady-state solutions are

$$f^{st}(\zeta_1, v_1, t) = exp[D_x\zeta_1 v_1 + \frac{D_x(B_z^2 + r_1)}{2}v_1^2], \tag{70}$$

$$f^{st}(\zeta_2, v_2, t) = exp[D_y\zeta_2 v_2 + \frac{D_y(B_z^2 + r_2)}{2}v_2^2]. \tag{71}$$

The corresponding Fourier-space solutions read

$$f(\zeta_1, v_1, t) = \Theta[t + v_1/[\zeta_1 - (B_z^2 + r_1)v_1]]f^{st}(\zeta_1, v_1, t), \tag{72}$$

$$f(\zeta_2, v_2, t) = \Theta[t + v_2/[(\zeta_2 - (B_z^2 + r_2)v_2]]f^{st}(\zeta_2, v_2, t). \tag{73}$$

After evaluation, we obtain

$$f(\zeta_1, v_1, t) = exp[-(D_x t + \frac{D_x \mathbf{B_z}^2 t^2}{2})\zeta_1^2 - \frac{D_x \mathbf{B_z}^4 t^2}{2}v_1^2], \tag{74}$$

$$f(\zeta_2, v_2, t) = exp[-(D_y t + \frac{D_y B_z^2 t^2}{2})\zeta_2^2 - \frac{D_y B_z^4 t^2}{2}v_2^2]. \tag{75}$$

For $D_x t \gg 1$ and $D_y t \gg 1$, the probability densities reduce to

$$f(x,t) = [4\pi D_x t]^{-\frac{1}{2}} exp[-\frac{x^2}{4D_x t}], \quad f(v_x,t) = [2\pi D_x B_z^4 t^2]^{-1/2} exp[-\frac{v_x^2}{2D_x B_z^4 t^2}], \tag{76}$$

$$f(y,t) = [4\pi D_y t]^{-\frac{1}{2}} exp[-\frac{y^2}{4D_y t}], \quad f(v_y,t) = [2\pi D_y B_z^4 t^2]^{-1/2} exp[-\frac{v_y^2}{2D_y B_z^4 t^2}]. \tag{77}$$

The corresponding mean-square values are

$$<x^2(t)> = 2D_x t, \quad <v_x^2(t)> = D_x B_z^4 t^2, \tag{78}$$

$$<y^2(t)> = 2D_y t, \quad <v_y^2(t)> = D_y B_z^4 t^2. \tag{79}$$

## 4. Vlasov equation with trap forces, thermal and active noises in the magnetic field

*4.1. Vlasov equation for a charged particle with trap forces and thermal noises*

The thermal fractional Vlasov equations in a uniform magnetic field $\vec{B} = B_z\vec{z}$ are modified as follows:

$$\frac{\partial}{\partial t}x = v_x + \xi_x^{th}(t), \tag{80}$$

$$\frac{\partial}{\partial t}v_x = -k_1 x - B_z^2 v_x - r_1 v_x, \tag{81}$$

and

$$\frac{\partial}{\partial t}y = v_y + \xi_y^{th}(t), \tag{82}$$

$$\frac{\partial}{\partial t}v_y = -k_2 y - B_z^{\ 2}v_y - r_2 v_y. \tag{83}$$

Here, the dimensionless parameters $m = 1$ and $q = 1$. The forces $-k_1 x$ and $-k_2 y$ represent harmonic trap forces. The thermal noises $\xi_x^{th}(t)$ and $\xi_y^{th}(t)$ are temporally correlated and satisfy $<\xi_j^{th}(t)\xi_j^{th}(t')> = D_j|\frac{t-t'}{\tau_{th}}|^{2h-2}$ for $j = x, y$, where the persistent Hurst exponent $h$ lies in the range $1/2 < h < 1$. The parameter $\tau_{th}$ is the thermal correlation time for a charged particle subject to trap forces and the thermal noise. The joint probability densities $f(x, v_x, t)$ and $f(y, v_y, t)$ are defined by $f(x, v_x, t) = <\delta(x - x(t))\delta(v_x - v_x(t))>$ and $f(y, v_y, t) = <\delta(y - y(t))\delta(v_y - v_y(t))>$. Taking time derivatives of the joint probability densities yields

$$\frac{\partial}{\partial t}f(x, v_x, t) = -\frac{\partial}{\partial x}<\frac{\partial x}{\partial t}\delta(x - x(t))\delta(v_x - v_x(t))> - \frac{\partial}{\partial v_x}<\frac{\partial v_x}{\partial t}\delta(x - x(t))\delta(v_x - v_x(t))>, \tag{84}$$

$$\frac{\partial}{\partial t}f(y, v_y, t) = -\frac{\partial}{\partial y}<\frac{\partial y}{\partial t}\delta(y - y(t))\delta\left(v_y - v_y(t)\right)> - \frac{\partial}{\partial v_y}<\frac{\partial v_y}{\partial t}\delta(y - y(t))\delta\left(v_y - v_y(t)\right)>. \tag{85}$$

Substituting Eqs. (80)-(81) into Eq. (84) and Eqs. (82)-(83) into Eq. (85), and following the procedure of Ref. [13], we obtain

$$\frac{\partial}{\partial t}f(x, v_x, t) = \left[-v_x\frac{\partial}{\partial x} + [k_1 x + (B_z^{\ 2} + r_1)v_x]\frac{\partial}{\partial v_x}\right]f(x, v_x, t) + D_x\left[\frac{t^{2h-1}}{(2h-1)\tau_{th}^{2h-1}}\frac{\partial^2}{\partial v_x \partial x} + \frac{t^{2h}}{2h\tau_{th}^{2h}}\frac{\partial^2}{\partial x^2}\right]f(x, v_x, t), \tag{86}$$

$$\frac{\partial}{\partial t}f(y, v_y, t) = \left[-v_y\frac{\partial}{\partial y} + [k_2 y + (B_z^{\ 2} + r_2)v_y]\frac{\partial}{\partial v_y}\right]f(y, v_y, t) + D_y\left[\frac{t^{2h-1}}{(2h-1)\tau_{th}^{2h-1}}\frac{\partial^2}{\partial v_y \partial y} + \frac{t^{2h}}{2h\tau_{th}^{2h}}\frac{\partial^2}{\partial y^2}\right]f(y, v_y, t). \tag{87}$$

The initial conditions are $x_0 = v_{x0} = 0$ and $y_0 = v_{y0} = 0$. Applying the double Fourier transform in Eq. (10), we obtain

$$\frac{\partial}{\partial t}f(\zeta_1, v_1, t) = -k_1 v_1\frac{\partial}{\partial \zeta_1} + \left[\zeta_1 - (B_z^{\ 2} + r_1)v_1\right]\frac{\partial}{\partial v_1}f(\zeta_1, v_1, t) - D_x\left[\frac{t^{2h-1}}{(2h-1)\tau_{th}^{2h-1}}\zeta_1 v_1 + \frac{t^{2h}}{2h\tau_{th}^{2h}}\zeta_1^2\right]f(\zeta_1, v_1, t), \tag{88}$$

$$\frac{\partial}{\partial t}f(\zeta_2, v_2, t) = -k_2 v_2\frac{\partial}{\partial \zeta_2} + \left[\zeta_2 - (B_z^{\ 2} + r_2)v_2\right]\frac{\partial}{\partial v_2}f(\zeta_2, v_2, t) - D_y\left[\frac{t^{2h-1}}{(2h-1)\tau_{th}^{2h-1}}\zeta_2 v_2 + \frac{t^{2h}}{2h\tau_{th}^{2h}}\zeta_2^2\right]f(\zeta_2, v_2, t). \tag{89}$$

Finally, we determine the probability densities for a charged particle with trap forces and exponentially correlated Gaussian forces in the regimes $t \ll \tau_{th}$, $t \gg \tau_{th}$, and for $\tau_{th} = 0$. The time domain $t \gg \tau_{th}$ is the one where the short-time singularity of the memory kernel averages out.

### 4.1.1. $f(x, t)$, $f(v_x, t)$ and $f(y, t)$, $f(v_y, t)$ in $t \ll \tau_{th}$

Since $\frac{\partial}{\partial t}f(\zeta_1, t) = 0$ and $\frac{\partial}{\partial t}f(v_1, t) = 0$, the Fourier transform of the probability densities $f(\zeta_1, t)$ and $f(v_1, t)$ reach the steady states, denoted by $f^{st}(\zeta_1, t)$ and $f^{st}(v_1, t)$, respectively. Accordingly, the two steady-state equations can be written as

$$-k_1 v_1 \frac{\partial}{\partial \zeta_1}f^{st}(\zeta_1, t) - \frac{D_x}{2}\left[\frac{t^{2h-1}}{(2h-1)\tau_{th}^{2h-1}}\zeta_1 v_1 + \frac{t^{2h}}{2h\tau_{th}^{2h}}\zeta_1^2\right] + C]f^{st}(\zeta_1, t), \tag{90}$$

$$\left[\zeta_1 - (B_z^{\ 2} + r_1)v_1\right]\frac{\partial}{\partial v_1}f^{st}(v_1, t) - \frac{D_x}{2}\left[\frac{t^{2h-1}}{(2h-1)\tau_{th}^{2h-2}}\zeta_1 v_1 + \frac{t^{2h}}{2h\tau_{th}^{2h-2}}\zeta_1^2\right] - C]f^{st}(v_1, t). \tag{91}$$

Here, $C$ is the separation constant. The Fourier transform of the probability density from Eq. (90) is given by

$$f^{st}(\zeta_1, t) = \exp\left[-\frac{D_x}{2k_1 v_1}\left[\frac{t^{2h-1}}{(2h-1)\tau_{th}^{2h-1}}\frac{\zeta_1^2}{2}v_1 + \frac{t^{2h}}{2h\tau_{th}^{2h}}\frac{\zeta_1^3}{3}\right] + \frac{C}{k_1 v_1}\zeta_1\right]. \tag{92}$$

To determine the Fourier transform of probability density for $\zeta_1$, we write $f(\zeta_1, t) = g(\zeta_1, t)f^{st}(\zeta_1, t)$. Retaining terms up to order $t^2/\tau_{th}^2$, the functions $g(\zeta_1, t)$ and $h(\zeta_1, t)$ are obtained as

$$g(\zeta_1, t) = h(\zeta_1, t)\exp\left[\frac{D_x}{2(k_1 v_1)^2}\left[\frac{t^{2h-2}}{\tau_{th}^{2h-1}}\frac{\zeta_1^3}{6}v_1 + \frac{t^{2h-1}}{\tau_{th}^{2h}}\frac{\zeta_1^4}{12}\right]\right], \tag{93}$$

$$h(\zeta_1, t) = p(\zeta_1, t)exp\left[-\frac{D_x}{2(k_1 v_1)^3}\left[\frac{(2h-1)t^{2h-2}}{\tau_{th}^{2h}}\frac{\zeta_1^5}{60}\right]\right]. \tag{94}$$

In Eq. (94), the function $p(\zeta_1, t)$ satisfies $\frac{\partial}{\partial t}p(\zeta_1, t) = -k_1 v_1 \frac{\partial}{\partial \zeta_1}p(\zeta_1, t)$. Since the Fourier transform of the probability density can be regarded as an arbitrary function of the variable $t - [\frac{\zeta_1}{k_1 v_1}]$, $p(\zeta_1, t)$, we obtain

$p(\zeta_1, t) = \Theta[t - [\frac{\zeta_1}{k_1 v_1}]]$. Keeping terms up to order $1/\tau_{th}^3$, the asymptotic form of the Fourier transform of the probability density becomes

$$f(\zeta_1, t) = p^{st}(\zeta_1, t)h^{st}(\zeta_1, t)g^{st}(\zeta_1, t)f^{st}(\zeta_1, t) = \Theta[t - [\frac{\zeta_1}{k_1 v_1}]]h^{st}(\zeta_1, t)g^{st}(\zeta_1, t)f^{st}(\zeta_1, t). \tag{95}$$

Following the same procedure, by using a similar procedure as in Eqs. (92)-(95), the Fourier transform $f(v_1, t)$ obtained from Eq. (91) is given by

$$f(v_1, t) = \Theta[t + v_1/[\zeta_1 - (B_z^2 + r_1)v_1]]h^{st}(v_1, t)g^{st}(v_1, t)f^{st}(v_1, t). \tag{96}$$

Using Eqs. (95) and (96), the joint Fourier-space probability density is obtained as

$$f(\zeta_1, v_1, t) = exp[-\frac{3D_x t^{2h+1}}{4\tau_{th}^{2h}}\zeta_1^2 - \frac{D_x k_1^2 t^{2h+3}}{24\tau_{th}^{2h}}v_1^2], \tag{97}$$

and, by an analogous procedure,

$$f(\zeta_2, v_2, t) = exp[-\frac{3D_y t^{2h+1}}{4\tau_{th}^{2h}}\zeta_2^2 - \frac{D_y k_2^2 t^{2h+3}}{24\tau_{th}^{2h}}v_2^2]. \tag{98}$$

Applying the inverse Fourier transform to Eqs. (97) and (98), the probability densities are obtained as

$$f(x, t) = \left[\pi \frac{3D_x t^{2h+1}}{\tau_{th}^{2h}}\right]^{-\frac{1}{2}} exp[-\frac{\tau_{th}^{2h} x^2}{3D_x t^{2h+1}}], \quad f(v_x, t) = \left[\pi \frac{D_y k_1^2 t^{2h+3}}{6\tau_{th}^{2h}}\right]^{-\frac{1}{2}} exp[-\frac{6\tau_{th}^{2h} v_x^2}{D_x k_1^2 t^{2h+3}}], \tag{99}$$

$$f(y, t) = \left[\pi \frac{3D_y t^{2h+1}}{\tau_{th}^{2h}}\right]^{-\frac{1}{2}} exp[-\frac{\tau_{th}^{2h-1} y^2}{3D_y t^{2h+1}}], \quad f(v_y, t) = \left[\pi \frac{D_y k_2^2 t^{2h+3}}{6\tau_{th}^{2h}}\right]^{-\frac{1}{2}} exp[-\frac{6\tau_{th}^{2h} v_y^2}{D_y k_2^2 t^{2h+3}}] \tag{100}$$

The mean-squared displacement and the mean-squared velocity are therefore given by

$$<x^2(t)> = \frac{3D_x}{2\tau_{th}^{2h}} t^{2h+1}, \quad <v_x^2(t)> = \frac{D_y k_1^2}{12\tau_{th}^{2h}} t^{2h+3}, \tag{101}$$

$$<y^2(t)> = \frac{3D_y}{2\tau_{th}^{2h}} t^{2h+1}, \quad <v_y^2(t)> = \frac{D_y k_2^2}{12\tau_{th}^{2h}} t^{2h+3}. \tag{102}$$

4.1.2. $f(x, t)$ and $f(v_x, t)$ in $t \gg \tau_{th}$

In this subsection, we derive the probability densities $f(x, t)$ and $f(v_x, t)$ in the long-time domain. From Eq. (88), the approximate equations for $f(\zeta_1, t)$ and $f(v_1, t)$ in the limit $t \gg \tau_{th}$ can be written as

$$\frac{\partial}{\partial t} f(\zeta_1, t) \cong -\frac{D_x}{2}[\frac{t^{2h-1}}{(2h-1)\tau_{th}^{2h-1}}\zeta_1 v_1 + \frac{t^{2h}}{2h\tau_{th}^{2h}}\zeta_1^2]f(\zeta_1, t), \tag{103}$$

$$\frac{\partial}{\partial t} f(v_1, t) \cong -\frac{D_x}{2}[\frac{t^{2h-1}}{(2h-1)\tau_{th}^{2h-1}}\zeta_1 v_1 + \frac{t^{2h}}{2h\tau_{th}^{2h}}\zeta_1^2]f(v_1, t), \tag{104}$$

From these equations, we obtain

$$f_{\zeta_1}(\zeta_1, t) = exp[-\frac{D_x}{2}\int[\frac{t^{2h-1}}{(2h-1)\tau_{th}^{2h-1}}\zeta_1 v_1 + \frac{t^{2h}}{2h\tau_{th}^{2h}}\zeta_1^2]dt], \tag{105}$$

$$f_{v_1}(v_1, t) = exp[-\frac{D_x}{2}\int[\frac{t^{2h-1}}{(2h-1)\tau_{th}^{2h-1}}\zeta_1 v_1 + \frac{t^{2h}}{2h\tau_{th}^{2h}}\zeta_1^2]dt]. \tag{106}$$

Writing $g_{\zeta_1}^{st}(\zeta_1, t)$ and $g_{v_1}^{st}(v_1, t)$ from $f(\zeta_1, v_1, t) = g_{\zeta_1}(\zeta_1, v_1, t)f_{\zeta_1}(\zeta_1, v_1, t)$, we obtain

$$g_{\zeta_1}^{st}(\zeta_1, t) = exp[\frac{D_x}{2}\int[\frac{t^{2h-1}}{(2h-1)\tau_{th}^{2h-1}}\zeta_1 v_1 + \frac{t^{2h}}{2h\tau_{th}^{2h}}\zeta_1^2]dt], \tag{107}$$

$$g_{v_1}^{st}(v_1, t) = exp[\frac{D_x}{2}\int[\frac{t^{2h-1}}{(2h-1)\tau_{th}^{2h-1}}\zeta_1 v_1 + \frac{t^{2h}}{2h\tau_{th}^{2h}}\zeta_1^2]dt]. \tag{108}$$

The steady-state Fourier transforms obtained from Eq. (92) are

$$f^{st}(\zeta_1, t) = \exp\left[-\frac{D_x}{2k_1 v_1}\left[\frac{t^{2h-1}}{(2h-1)\tau_{th}^{2h-1}}\frac{\zeta_1^2}{2}v_1 + \frac{t^{2h}}{2h\tau_{th}^{2h}}\frac{\zeta_1^3}{3}\right] + \frac{C}{k_1 v_1}\zeta_1\right], \tag{109}$$

and

$$f^{st}(v_1,t) = \exp\left[\frac{D_x}{2[\zeta_1-(B_z{}^2+r_1)v_1]}\left[\frac{t^{2h-1}}{(2h-1)\tau_{th}^{2h-1}}\zeta_1\frac{v_1^2}{2}+\frac{t^{2h}}{2h\tau_{th}^{2h}}\zeta_1^2 v_1\right]+\frac{C}{k_1 v_1}v_1\right]. \tag{110}$$

Thus

$$g(\zeta_1,t) = h(\zeta_1,t)g_{\zeta_1}^{st}(\zeta_1,t) = h(\zeta_1,t)\,exp\left[\frac{D_x}{2}\int\left[\frac{t^{2h-1}}{(2h-1)\tau_{th}^{2h-1}}\zeta_1 v_1+\frac{t^{2h}}{2h\tau_{th}^{2h}}\zeta_1^2\right]dt\right], \tag{111}$$

$$g(v_1,t) = h(v_1,t)g_{\zeta_2}^{st}(v_1,t) = h(v_1,t)\,exp\left[\frac{D_x}{2}\int\left[\frac{t^{2h-1}}{(2h-1)\tau_{th}^{2h-1}}\zeta_1 v_1+\frac{t^{2h}}{2h\tau_{th}^{2h}}\zeta_1^2\right]dt\right]. \tag{112}$$

Taking the solutions as arbitrary functions of variables $t-\zeta_1/[(k_1 v_1]$ and $t+v_1/[(\zeta_1 - B_z{}^2 v_1]$, we obtain $h(\zeta_1,v_1,t) = \Theta[t-\zeta_1/[(k_1 v_1]]$ and $h(\zeta_2,v_2,t) = \Theta[t+v_1/[\zeta_1 - B_z{}^2 v_1]]$. After expanding in powers of $t/\tau_t$ and performing cancellations, we obtain

$$f(\zeta_1,t) = \Theta[t-\zeta_1/[(k_1 v_1]]g_{\zeta_1}^{st}(\zeta_1,t)f^{st}(\zeta_1,t), \tag{113}$$

$$f(v_1,t) = \Theta[t+v_1/[\zeta_1-(B_z{}^2+r_1)v_1]]g_{v_1}^{st}(v_1,t)f^{st}(v_1,t). \tag{114}$$

Hence,

$$f(\zeta_1,v_1,t)=f(\zeta_1,t)\,f(v_1,t) = exp\left[-\frac{D_x t^{2h+1}}{4h(2h+1)\tau_{th}^{2h-1}}\zeta_1^2 - \frac{D_x k_1^2 t^{2h+3}}{4h(2h+1)\tau_{th}^{2h}}v_1^2\right]. \tag{115}$$

Applying the inverse Fourier transform yields

$$f(x,t) = \left[\pi\frac{D_x t^{2h+1}}{h(2h+1)\tau_{th}^{2h-1}}\right]^{-\frac{1}{2}}exp\left[-\frac{h(2h+1)\tau_{th}^{2h-1}x^2}{D_x t^{2h+1}}\right],\; f(v_x,t) = \left[\pi\frac{D_x k_1^2 t^{2h+3}}{h(2h+1)\tau_{th}^{2h}}\right]^{-\frac{1}{2}}exp\left[-\frac{h(2h+1)\tau_{th}^{2h}v_x^2}{D_x k_1^2 t^{2h+3}}\right]. \tag{116}$$

The mean squared displacement and velocity are therefore

$$<x^2(t)> = \frac{D_x}{2h(2h+1)\tau_{th}^{2h-1}}t^{2h+1},\;\; <v_x^2(t)> = \frac{D_x k_1^2}{2h(2h+1)\tau_{th}^{2h}}t^{2h+3}. \tag{117}$$

We write approximate equations for the time derivatives of $f(\zeta_2,t)$ and $f(v_2,t)$ from Eq. (89) in the long-time domain as

$$\frac{\partial}{\partial t}f(\zeta_2,t) \cong \frac{D_y}{2}\left[\frac{t^{2h-1}}{(2h-1)\tau_{th}^{2h-2}}\zeta_2 v_2+\frac{t^{2h}}{2h\tau_{th}^{2h-2}}\zeta_2^2\right]f(\zeta_2,t), \tag{118}$$

$$\frac{\partial}{\partial t}f(v_2,t) \cong \frac{D_y}{2}\left[\frac{t^{2h-1}}{(2h-1)\tau_{th}^{2h-2}}\zeta_2 v_2+\frac{t^{2h}}{2h\tau_{th}^{2h-2}}\zeta_2^2\right]f(v_2,t), \tag{119}$$

By using a similar procedure of Eq. (105)-(114) for $\zeta_2$ and $v_2$, we get the Fourier transforms of probability densities $f(\zeta_2,t)$ and $f(v_2,t)$ as

$$f(\zeta_2,t) = h(\zeta_2,t)g_{\zeta_1}^{st}(\zeta_2,t)f^{st}(\zeta_2,t) = \Theta[t-\zeta_2/[k_2 v_2]]g_{\zeta_1}^{st}(\zeta_2,t)f^{st}(\zeta_2,t), \tag{120}$$

$$f(v_2,t) = h(v_2,t)g_{v_2}^{st}(v_2,t)f^{st}(v_2,t) = \Theta[t+v_2/[\zeta_2-(B_z{}^2+r_2)v_2]]g_{v_2}^{st}(v_2,t)f^{st}(v_2,t). \tag{121}$$

From Eq. (120) and Eq. (121), we calculate $f(\zeta_2,v_2,t)= f(\zeta_2,t)\,f(v_2,t)$ as

$$f(\zeta_2,v_2,t)=f(\zeta_2,t)\,f(v_2,t) = exp\left[-\frac{D_x t^{2h+1}}{4h(2h+1)\tau_{th}^{2h-1}}\zeta_2^2 - \frac{D_x k_1^2 t^{2h+3}}{4h(2h+1)\tau_{th}^{2h}}v_2^2\right]. \tag{122}$$

We get the probability densities $f(y,t)$ and $f(v_y,t)$ from Eq. (122) as

$$f(y,t) = \left[\pi\frac{D_x t^{2h+1}}{h(2h+1)\tau_{th}^{2h-1}}\right]^{-\frac{1}{2}}exp\left[-\frac{h(2h+1)\tau_{th}^{2h-1}y^2}{D_y t^{2h+1}}\right],\; f(v_y,t) = \left[\pi\frac{D_y k_2^2 t^{2h+3}}{h(2h+1)\tau_{th}^{2h}}\right]^{-\frac{1}{2}}exp\left[-\frac{h(2h+1)\tau_{th}^{2h}v_y^2}{D_y k_2^2 t^{2h+3}}\right], \tag{123}$$

with the mean squared displacement and the mean squared velocity for $f(y,t)$ and $f(v_y,t)$

$$<y^2(t)> = \frac{D_x}{2h(2h+1)\tau_{th}^{2h-1}}t^{2h+1},\;\; <v_y^2(t)> = \frac{D_y k_2^2}{2h(2h+1)\tau_{th}^{2h}}t^{2h+3}. \tag{124}$$

4.1.3. $f(x,t)$, $f(v_x,t)$ and $f(y,t)$, $f(v_y,t)$ in $\tau_{th}=0$

We write the approximate equation from Eq. (88) for $\zeta_1$, $v_1$ and Eq. (89) for $\zeta_2$, $v_2$ as

$$\frac{\partial}{\partial t}f(\zeta_1,t) \cong -k_1 v_1 \frac{\partial}{\partial \zeta_1} f(\zeta_1,t) - \frac{D_x}{2}\left[\frac{t^{2h}}{2h\tau_{th}^{2h-2}}\right]\zeta_1^2 f(\zeta_1,t), \tag{125}$$

$$\frac{\partial}{\partial t}f(v_1,t) \cong [\zeta_1 - (B_z^2+r_1)v_1]\frac{\partial}{\partial v_1} f(v_1,t) - \frac{D_x}{2}\left[\frac{t^{2h}}{2h\tau_{th}^{2h}}\right]\zeta_1^2 f(v_1,t), \tag{126}$$

$$\frac{\partial}{\partial t}f(\zeta_2,t) \cong -k_2 v_2 \frac{\partial}{\partial \zeta_2} f(\zeta_2,t) - \frac{D_y}{2}\left[\frac{t^{2h}}{2h\tau_{th}^{2h}}\right]\zeta_2^2 f(\zeta_2,t), \tag{127}$$

$$\frac{\partial}{\partial t}f(v_2,t) \cong [\zeta_2 - (B_z^2+r_2)v_2]\frac{\partial}{\partial v_2} f(v_2,t) - \frac{D_y}{2}\left[\frac{t^{2h}}{2h\tau_{th}^{2h}}\right]\zeta_2^2 f(v_2,t). \tag{128}$$

From the above equations in the steady state, we calculate $f^{st}(\zeta_1,t)$, $f^{st}(v_1,t)$ and $f^{st}(\zeta_2,t)$, $f^{st}(v_2,t)$ as

$$f^{st}(\zeta_1,t) = exp\left[-\frac{D_x}{2k_1 v_1}\left[\frac{t^{2h}}{2h\tau_{th}^{2h}}\right]\frac{\zeta_1^3}{3}\right], \quad f^{st}(v_1,t) = exp\left[\frac{D_x}{2[\zeta_1 - B_z^2 v_1]}\left[\frac{t^{2h}}{2h\tau_{th}^{2h}}\right]\zeta_1^2 v_1\right], \tag{129}$$

$$f^{st}(\zeta_2,t) = exp\left[-\frac{D_y}{2k_2 v_2}\left[\frac{t^{2h}}{2h\tau_{th}^{2h}}\right]\frac{\zeta_2^3}{3}\right], \quad f^{st}(v_2,t) = exp\left[\frac{D_y}{2[\zeta_2 - B_z^2 v_2]}\left[\frac{t^{2h}}{2h\tau_{th}^{2h}}\right]\zeta_2^2 v_2\right]. \tag{130}$$

The Fourier transforms of the probability densities $f(\zeta_1,t), f(v_1,t)$ and $f(\zeta_2,t), f(v_2,t)$ are derived as

$$f(\zeta_1,t) = \Theta[t - \zeta_1/k_1 v_1]f^{st}(\zeta_1,t), \quad f(v_1,t) = \Theta[t + v_1/[(\zeta_1 - (B_z^2+r_1)v_1]]f^{st}(v_1,t), \tag{131}$$

$$f(\zeta_2,t) = \Theta[t - \zeta_2/k_2 v_2]f^{st}(\zeta_2,t), \quad f(v_2,t) = \Theta[t + v_2/[(\zeta_2 - (B_z^2 + r_2)v_2]]f^{st}(v_2,t). \tag{132}$$

We calculate $f(\zeta_1,v_1,t)$ and $f(\zeta_2,v_2,t)$ from Eqs. (131) and (132) as

$$f(\zeta_1,v_1,t) = f(\zeta_1,t)f(v_1,t) = exp\left[-\frac{D_x t^{2h+1}}{2h\tau_{th}^{2h}}\zeta_1^2 - \frac{D_x k_1^2 t^{2h+3}}{12h\tau_{th}^{2h}}v_1^2\right], \tag{133}$$

$$f(\zeta_2,v_2,t) = f(\zeta_2,t)f(v_2,t) = exp\left[-\frac{D_y t^{2h+1}}{2h\tau_{th}^{2h}}\zeta_2^2 - \frac{D_y k_2^2 t^{2h+3}}{12h\tau_{th}^{2h}}v_2^2\right]. \tag{134}$$

By using the inverse Fourier transforms, $f(x,t), f(v_x,t)$ and $f(y,t), f(v_y,t)$ are, respectively, presented by

$$f(x,t) = \left[2\pi \frac{D_x t^{2h+1}}{h\tau_{th}^{2h}}\right]^{-\frac{1}{2}} exp\left[-\frac{h\tau_{th}^{2h}x^2}{2D_x t^{2h+1}}\right], \quad f(v_x,t) = \left[\pi \frac{D_x k_1^2 t^{2h+3}}{3h\tau_{th}^{2h}}\right]^{-\frac{1}{2}} exp\left[-\frac{3h\tau_{th}^{2h}v_x^2}{D_x k_1^2 t^{2h+3}}\right], \tag{135}$$

$$f(y,t) = \left[2\pi \frac{D_y B_z^2 t^{2h+1}}{h\tau_{th}^{2h}}\right]^{-\frac{1}{2}} exp\left[-\frac{h\tau_{th}^{2h}y^2}{2D_y t^{2h+1}}\right], \quad f(v_y,t) = \left[\pi \frac{D_y k_2^2 t^{2h+3}}{3h\tau_{th}^{2h}}\right]^{-\frac{1}{2}} exp\left[-\frac{3h\tau_{th}^{2h}v_y^2}{D_y k_2^2 t^{2h+3}}\right]. \tag{136}$$

The mean squared displacement and the mean squared velocity for $f(x,t), f(v_x,t)$ and $f(y,t), f(v_y,t)$ are, respectively, calculated as

$$<x^2(t)> = \frac{D_x}{h\tau_{th}^{2h}}t^{2h+1}, \quad <v_x^2(t)> = \frac{D_x k_1^2}{6h\tau_{th}^{2h}}t^{2h+3}, \tag{137}$$

$$<y^2(t)> = \frac{D_y}{h\tau_{th}^{2h}}t^{2h+1}, \quad <v_y^2(t)> = \frac{D_y k_2^2}{6h\tau_{th}^{2h}}t^{2h+3}. \tag{138}$$

*4.2. Vlasov equation for a charged particle with trap forces and active noises*

The Vlasov equations of motion for a charged particle with trap forces and active noises $\xi_x^{ac}(t)$, $\xi_y^{ac}(t)$ in the magnetic field $\vec{B} = B_z \vec{z}$ are given by

$$\frac{\partial}{\partial t}x = v_x + \xi_x^{ac}(t), \tag{139}$$

$$m\frac{\partial}{\partial t}v_x = -k_1 x - B_z^2 v_x - \gamma_1 \int_0^t dt' \left|\frac{t-t'}{\tau_{th}}\right|^{2h-2} v_x(t') - r_1 v_x, \tag{140}$$

and

$$\frac{\partial}{\partial t} y = v_y + \xi_y^{ac}(t), \qquad (141)$$

$$m\frac{\partial}{\partial t} v_y = -k_2 y - B_z^{\,2} v_y - \gamma_2 \int_0^t dt' \left|\frac{t-t'}{\tau_{th}}\right|^{2h-2} v_y(t') - r_2 v_y. \qquad (142)$$

Here, $-k_1 x$ and $-k_2 y$ denote the trap forces. The active noises $\xi_x^{ac}(t)$ and $\xi_x^{ac}(t)$ depends on the time difference: $<\xi_j^{ac}(t)\xi_j^{ac}(t')> = \frac{D_i}{\tau_{ac}} exp(\frac{|t-t'|}{\tau_{ac}})$ for $j = x, y$ and $\tau_{ac}$ is the active correlation time for a charged particle with the trap force and the active noise.

The joint probability densities $p(x, v_x, t)$ and $p(y, v_y, t)$ are defined by $f(x, v_x, t) = <\delta(x - x(t))\delta(v_x - v_x(t))>$ and $f(y, v_y, t) = <\delta(y - y(t))\delta(v_y - v_y(t))>$. By taking time derivatives of the joint probability density, we have the differential equations as

$$\frac{\partial}{\partial t} f(x, v_x, t) = -\frac{\partial}{\partial x} <\frac{\partial x}{\partial t}\delta(x - x(t))\delta(v_x - v_x(t))> - \frac{\partial}{\partial v_x} <\frac{\partial v_x}{\partial t}\delta(x - x(t))\delta(v_x - v_x(t))>, \qquad (143)$$

$$\frac{\partial}{\partial t} f(y, v_y, t) = -\frac{\partial}{\partial y} <\frac{\partial y}{\partial t}\delta(y - y(t))\delta(v_y - v_y(t))> - \frac{\partial}{\partial v_y} <\frac{\partial v_y}{\partial t}\delta(y - y(t))\delta(v_y - v_y(t))>. \qquad (144)$$

Inserting Eqs. (139) and (140) into Eq. (143) and inserting Eqs. (141) and (142) into Eq. (144), we have, manipulating the method of Ref. [13],

$$\frac{\partial}{\partial t} f(x, v_x, t) = \left[-v_x\frac{\partial}{\partial x} + [k_1 x + (B_z^{\,2} + r_1)v_x]\frac{\partial}{\partial v_x}\right] f(x, v_x, t) + D_x[-b_1(t)\frac{\partial^2}{\partial x^2} + a_1(t)\frac{\partial^2}{\partial v_x^2}] f(x, v_x, t), \qquad (145)$$

$$\frac{\partial}{\partial t} f(y, v_y, t) = \left[-v_y\frac{\partial}{\partial y} + [k_2 y + (B_z^{\,2} + r_2)v_y]\frac{\partial}{\partial v_y}\right] f(y, v_y, t) + D_y[-b_2(t)\frac{\partial^2}{\partial y^2} + a_2(t)\frac{\partial^2}{\partial v_y^2}] f(y, v_y, t). \qquad (146)$$

where the dimensionless $m = 1$ and $q = 1$. The parameters $a_1(t)$, $a_2(t)$, $b_1(t)$, and $b_2(t)$ are, respectively, given by $a_1(t) = a_2(t) = 1 - exp(-t/\tau)$, and $b_1(t) = b_2(t) = (t + \tau) exp(-t/\tau) - \tau$. By using the double Fourier transform of Eq. (10), we derive the Fourier transforms of Eq. (87) and Eq. (88) as

$$\frac{\partial}{\partial t} f(\zeta_1, v_1, t) = -k_1 v_1 \frac{\partial}{\partial \zeta_1} + [\zeta_1 - (B_z^{\,2} + r_1)v_1]\frac{\partial}{\partial v_1} f(\zeta_1, v_1, t) + D_x[b_1(t)\zeta_1 v_1 - a_1(t)\zeta_1^2] f(\zeta_1, v_1, t), \qquad (147)$$

$$\frac{\partial}{\partial t} f(\zeta_2, v_2, t) = -k_2 v_2 \frac{\partial}{\partial \zeta_2} + [\zeta_2 - (B_z^{\,2} + r_2)v_2]\frac{\partial}{\partial v_2} f(\zeta_2, v_2, t) + D_y[b_2(t)\zeta_2 v_2 - a_2(t)\zeta_2^2] f(\zeta_2, v_2, t). \qquad (148)$$

Here, the initial condition is $x_0 = v_{x0} = 0$, $y_0 = v_{y0} = 0$.

Now, we find the Probability Densities for a charged particle with trap forces and exponential correlated Gaussian forces in $t \ll \tau_{ac}$, $t \gg \tau_{ac}$, and for $\tau_{ac} = 0$. The limit domain $t \gg \tau_{ac}$ is the one where the active noise is reduced to effective white noise.

4.2.1. $f(x,t)$, $f(v_x,t)$ and $f(y,t)$, $f(v_y,t)$ in $t \ll \tau_{ac}$

As we have $\frac{\partial}{\partial t}f(\zeta_1,t) = 0$ and $\frac{\partial}{\partial t}f(v_1,t) = 0$, the Fourier transform of the probability density $f(\zeta_1, t)$ and $f(v_1, t)$ in the steady state becomes $f^{st}(\zeta_1, t)$ and $f^{st}(v_1, t)$. We write the two steady equations in the form:

$$-k_1 v_1 \frac{\partial}{\partial \zeta_1} f^{st}(\zeta_1, t) + \frac{D_x}{2}[b_1(t)\zeta_1 v_1 - a_1(t)\zeta_1^2] + E] f^{st}(\zeta_1, t), \qquad (149)$$

$$[\zeta_1 - (B_z^{\,2} + r_1)v_1]\frac{\partial}{\partial v_1} f^{st}(v_1, t) + \frac{D_x}{2}[b_1(t)\zeta_1 v_1 - a_1(t)\zeta_1^2] - E] f^{st}(v_1, t). \qquad (150)$$

Here, $E$ is the separation constant. The Fourier transform of the probability density from Eq. (149) is calculated as

$$f^{st}(\zeta_1, t) = exp\left[\frac{D_x}{2k_1 v_1}\left[b_1(t)\frac{\zeta_1^2}{2}v_1 - a_1(t)\frac{\zeta_1^3}{3}\right] + \frac{E}{k_1 v_1}\zeta_1\right]. \qquad (151)$$

In order to find the Fourier transform of probability density for $\zeta_1$ from $f(\zeta_1, t) = g(\zeta_1, t)f^{st}(\zeta_1, t)$, the Fourier transform of the probability density via the calculation including terms up to order $t^2/\tau^2$ are calculated as

$$g(\zeta_1, t) = h(\zeta_1, t) exp\left[-\frac{D_x}{2(k_1 v_1)^2}\left[b_1'(t)\frac{\zeta_1^3}{6}v_1 - a_1'(t)\frac{\zeta_1^4}{12}\right]\right], \qquad (152)$$

$$h(\zeta_1, t) = p(\zeta_1, t) exp\left[\frac{D_x}{2(k_1 v_1)^3}\left[b_1''(t)\frac{\zeta_1^4}{24}v_1 - a_1''(t)\frac{\zeta_1^5}{60}\right]\right], \qquad (153)$$

$$p(\zeta_1, t) = q(\zeta_1, t) exp\left[-\frac{D_x}{2(k_1 v_1)^4}\left[b_1'''(t)\frac{\zeta_1^6}{120}v_1\right]\right]. \qquad (154)$$

In Eq. (88), $q(\zeta_1,t)$ obeys $\frac{\partial}{\partial t}q(\zeta_1,t) = -k_1 v_1 \frac{\partial}{\partial \zeta_1} q(\zeta_1,t)$. As we take the Fourier transform of the probability density as an arbitrary function of the variable $t - [\frac{\zeta_1}{k_1 v_1}]$, $q(\zeta_1,t)$ is given by $q(\zeta_1,t) = \Theta[t - [\frac{\zeta_1}{k_1 v_1}]]$. The asymptotic Fourier transform of the probability density cut off in $q(\zeta_1,t)$ up to the terms proportional to $1/\tau_{ac}^3$. We consequently get $f(\zeta_1,t)$ related to $q(\zeta_1,t)$ as

$$f(\zeta_1,t) = q^{st}(\zeta_1,t)p^{st}(\zeta_1,t)h^{st}(\zeta_1,t)g^{st}(\zeta_1,t)f^{st}(\zeta_1,t) = \Theta[t - [\frac{\zeta_1}{k_1 v_1}]]p^{st}(\zeta_1,t)h^{st}(\zeta_1,t)g^{st}(\zeta_1,t)f^{st}(\zeta_1,t). \quad (155)$$

By using a similar procedure of Eq. (151)-(155) for $\zeta_1$, $f(v_1,t)$ is calculated as

$$f(v_1,t) = \Theta[t + v_1/[\zeta_1 - (B_z^2 + r_1)v_1]]p^{st}(v_1,t)h^{st}(v_1,t)g^{st}(v_1,t)f^{st}(v_1,t). \quad (156)$$

we calculate $f(\zeta_1,v_1,t) = f(\zeta_1,t)f(v_1,t)$ from Eqs. (155) and (156) as

$$f(\zeta_1,v_1,t) = exp[-\frac{D_x t^3}{4k_1}\zeta_1^2 - \frac{D_x B_z^4 t^3}{4k_1}v_1^2 - \frac{D_x t^3}{4}\zeta_1^2 - \frac{D_x B_z^4 t^3}{4}v_1^2]. \quad (157)$$

From Eq. (157), using the inverse Fourier transform, the probability densities $f(x,t)$ and $f(v_x,t)$ are, respectively, presented by

$$f(x,t) = [\pi D_x k_1^{-1} t^3]^{-\frac{1}{2}} exp[-\frac{k_1 x^2}{D_x t^3}], \quad f(v_x,t) = [\pi D_x B_z^4 k_1^{-1} t^3]^{-\frac{1}{2}} exp[-\frac{k_1 x^2}{D_x B_z^4 t^3}]. \quad (158)$$

The mean squared displacement and the mean squared velocity for $f(x,t)$ and $f(v_x,t)$ are, respectively, given by

$$<x^2(t)> = \frac{D_x}{2k_1}t^3, \quad <v_x^2(t)> = \frac{D_x B_z^4}{2k_1}t^3. \quad (159)$$

From Eq. (148), using a similar procedure of Eq. (149)-(157) for $\zeta_1$ and $v_1$, the probability densities $f(y,t)$ and $f(v_y,t)$ are, respectively, presented by

$$f(y,t) = [\pi D_y k_2^{-1} t^3]^{-\frac{1}{2}} exp[-\frac{k_2 y^2}{D_y t^3}], \quad f(v_y,t) = [\pi D_y B_z^4 k_2^{-1} t^3]^{-\frac{1}{2}} exp[-\frac{k_2 v_y^2}{D_y B_z^4 t^3}]. \quad (158)$$

The mean squared displacement and the mean squared velocity for $f(x,t)$ and $f(v_x,t)$ are, respectively, given by

$$<y^2(t)> = \frac{D_y}{2k_2}t^3, \quad <v_y^2(t)> = \frac{D_y B_z^4}{2k_2}t^3. \quad (159)$$

4.2.2. $f(x,t)$ and $f(v_x,t)$ in $t \gg \tau_{ac}$

In this subsection, we find the probability densities in the long-time domain $t \gg \tau_{ac}$. We can write approximate equations for $f(\zeta_1,t)$ and $f(v_1,t)$ from Eq. (147) in the long-time domain as

$$\frac{\partial}{\partial t}f(\zeta_1,t) \cong \frac{D_x}{2}[b_1(t)\zeta_1 v_1 - a_1(t)\zeta_1^2]f(\zeta_1,t), \quad (160)$$

$$\frac{\partial}{\partial t}f(v_1,t) \cong \frac{D_x}{2}[b_1(t)\zeta_1 v_1 - a_1(t)\zeta_1^2]f(v_1,t), \quad (161)$$

From the above equations, we have

$$f_{\zeta_1}(\zeta_1,t) = exp[\frac{D_x}{2}\int[b_1(t)\zeta_1 v_1 - a_1(t)\zeta_1^2]dt], \quad (162)$$

$$f_{v_1}(v_1,t) = exp[\frac{D_x}{2}\int[b_1(t)\zeta_1 v_1 - a_1(t)\zeta_1^2]dt. \quad (163)$$

We get $g_{\zeta_1}^{st}(\zeta_1,t)$ and $g_{v_1}^{st}(v_1,t)$ from $f(\zeta_1,v_1,t) = g_{\zeta_1}(\zeta_1,v_1,t)f_{\zeta_1}(\zeta_1,v_1,t)$ and $f(v_1,t) = g_{v_1}(v_1,t)f_{v_1}(v_1,t)$ as

$$g_{\zeta_1}^{st}(\zeta_1,t) = exp[-\frac{D_x}{2}\int[b_1(t)\zeta_1 v_1 - a_1(t)\zeta_1^2]dt], \quad (164)$$

$$g_{v_1}^{st}(v_1,t) = exp[-\frac{D_x}{2}\int[b_1(t)\zeta_1 v_1 - a_1(t)\zeta_1^2]dt]. \quad (165)$$

$f^{st}(\zeta_1,t)$ from Eq. (151) is given by

$$f^{st}(\zeta_1, t) = \exp\left[\frac{D_x}{2k_1 v_1}\left[b_1(t)\frac{\zeta_1^2}{2}v_1 - a_1(t)\frac{\zeta_1^3}{3}\right] + \frac{C}{k_1 v_1}\zeta_1\right], \tag{166}$$

and $f^{st}(v_1, t)$ is calculated as

$$f^{st}(v_1, t) = exp\left[\frac{D_x}{\zeta_1}[b_1(t)\zeta_1 \frac{v_1^2}{2} - a_1(t)\zeta_1^2 v_1] + \frac{D_x B_z^2}{\zeta_1^2}[b_1(t)\zeta_1 \frac{v_1^3}{3} - a_1(t)\zeta_1^2 \frac{v_1^2}{2}]\right]. \tag{167}$$

The Fourier transforms of the probability densities $g(\zeta_1, v_1, t)$ and $g(v_1, t)$ are given by

$$g(\zeta_1, t) = h(\zeta_1, t) g_{\zeta_1}^{st}(\zeta_1, t) = h(\zeta_1, t) \exp\left[-\frac{D_x}{2}\int[b_1(t)\zeta_1 v_1 - a_1(t)\zeta_1^2]dt\right], \tag{168}$$

$$g(v_1, t) = h(v_1, t) g_{\zeta_2}^{st}(v_1, t) = h(v_1, t) \exp\left[-\frac{D_x}{2}\int[b_1(t)\zeta_1 v_1 - a_1(t)\zeta_1^2]dt\right]. \tag{169}$$

Here, $\int a_i(t)dt = t - \tau_{ac}$, $a_i(t) = 1$ and $\int b_i(t)dt = -\tau_{ac} t$, $b_i(t) = -\tau_{ac}$ for $i = 1,2$ in the long-time domain. Taking the solutions as arbitrary functions of variables $t - \zeta_1/[(k_1 v_1]$ and $t + v_1/[(\zeta_1 - B_z^2 v_1]$, the arbitrary functions $h(\zeta_1, v_1, t)$ and $h(\zeta_2, v_2, t)$ become $h(\zeta_1, v_1, t) = \Theta[t - \zeta_1/[(k_1 v_1]]$ and $h(\zeta_2, v_2, t) = \Theta[t + v_1/[\zeta_1 - B_z^2 v_1]]$. By expanding in powers of $t/\tau_t$, we derive and calculate the expression for $f(\zeta_1, t)$ and $f(v_1, t)$ after some cancellations, as

$$f(\zeta_1, t) = h(\zeta_1, t) g_{\zeta_1}^{st}(\zeta_1, t) f^{st}(\zeta_1, t) = \Theta[t - \zeta_1/[(k_1 v_1]] g_{\zeta_1}^{st}(\zeta_1, t) f^{st}(\zeta_1, t), \tag{170}$$

$$f(v_1, t) = h(v_1, t) g_{v_1}^{st}(v_1, t) f^{st}(v_1, t) = \Theta[t + v_1/[\zeta_1 - (B_z^2 + r_1)v_1]] g_{v_1}^{st}(v_1, t) f^{st}(v_1, t). \tag{171}$$

Therefore, from Eq. (170) and Eq. (171), we calculate $f(\zeta_1, v_1, t) = f(\zeta_1, t) f(v_1, t)$ as

$$f(\zeta_1, v_1, t) = f(\zeta_1, t) f(v_1, t) = exp\left[-\frac{D_x t^3}{4k_1}\zeta_1^2 - \frac{D_x B_z^4 t^3}{4k_1}v_1^2 - \frac{D_x t^3}{4}\zeta_1^2 - \frac{D_x B_z^4 t^3}{4}v_1^2\right]. \tag{172}$$

From Eqs. (172), using the inverse Fourier transform, the probability densities $f(x, t)$ and $f(v_x, t)$ are, respectively, presented by

$$f(x, t) = \left[\pi D_x k_1^{-1} t^3\right]^{-\frac{1}{2}} exp\left[-\frac{k_1 x^2}{D_x t^3}\right], \quad f(v_x, t) = \left[\pi D_x B_z^4 k_1^{-1} t^3\right]^{-\frac{1}{2}} exp\left[-\frac{k_1 v_x^2}{D_x B_z^4 t^3}\right]. \tag{173}$$

The mean squared displacement and the mean squared velocity for $f(x, t)$ and $f(v_x, t)$ are, respectively, given by

$$<x^2(t)> = \frac{D_x}{2k_1}t^3, \quad <v_x^2(t)> = \frac{D_x B_z^4}{2k_1}t^3. \tag{174}$$

In $t \gg \tau_{ac}$, we write approximate equations for the time derivatives of $f(\zeta_2, t)$ and $f(v_2, t)$ from Eq. (148) in the long-time domain as

$$\frac{\partial}{\partial t}f(\zeta_2, t) \cong \frac{D_y}{2}[b_2(t)\zeta_2 v_2 - a_2(t)\zeta_2^2]f(\zeta_2, t), \tag{175}$$

$$\frac{\partial}{\partial t}f(v_2, t) \cong \frac{D_y}{2}[b_2(t)\zeta_2 v_2 - a_2(t)\zeta_2^2]f(v_2, t), \tag{176}$$

By using a similar procedure of Eq. (162)-(171) for $\zeta_2$ and $v_2$, we get the Fourier transforms of probability densities $f(\zeta_2, t)$ and $f(v_2, t)$ as

$$f(\zeta_2, t) = h(\zeta_2, t) g_{\zeta_1}^{st}(\zeta_2, t) f^{st}(\zeta_2, t) = \Theta[t - \zeta_2/[k_2 v_2]] g_{\zeta_1}^{st}(\zeta_2, t) f^{st}(\zeta_2, t), \tag{177}$$

$$f(v_2, t) = h(v_2, t) g_{v_2}^{st}(v_2, t) f^{st}(v_2, t) = \Theta[t + v_2/[\zeta_2 - (B_z^2 + r_2)v_2]] g_{v_2}^{st}(v_2, t) f^{st}(v_2, t). \tag{178}$$

From Eq. (167) and Eq. (168), we calculate $f(\zeta_2, v_2, t) = f(\zeta_2, t) f(v_2, t)$ as

$$f(\zeta_2, v_2, t) = f(\zeta_2, t) f(v_2, t) = exp\left[-\frac{D_y t^3}{4k_2}\zeta_2^2 - \frac{D_y B_z^4 t^3}{4k_2}v_2^2 - \frac{D_y t^3}{4}\zeta_2^2 - \frac{D_y B_z^4 t^3}{4}v_2^2\right]. \tag{179}$$

We get the probability densities $f(y, t)$ and $f(v_y, t)$ from Eq. (123) as

$$f(y, t) = \left[\pi D_y k_2^{-1} t^3\right]^{-\frac{1}{2}} exp\left[-\frac{k_2 y^2}{D_y t^3}\right], \quad f(v_y, t) = \left[\pi D_y B_z^4 k_2^{-1} t^3\right]^{-\frac{1}{2}} exp\left[-\frac{k_2 v_y^2}{D_y B_z^4 t^3}\right], \tag{180}$$

with the mean squared displacement and the mean squared velocity for $f(y,t)$ and $f(v_y,t)$

$$< y^2(t) > = \frac{D_y}{2k_2} t^3, \quad < v_y^2(t) > = \frac{D_y B_z^4}{2k_2} t^3. \tag{181}$$

4.2.3. $f(x,t)$, $f(v_x,t)$ and $f(y,t)$, $f(v_y,t)$ in $\tau_{ac} = 0$

For $\tau_{ac} = 0$ $(a_1(t) = a_2(t) = 1$ and $b_1(t) = b_2(t) = 0)$, we write the approximate equation from Eq. (147) for $\zeta_1$, $v_1$ and Eq. (138) for $\zeta_2$, $v_2$ as

$$\frac{\partial}{\partial t} f(\zeta_1, t) \cong -k_1 v_1 \frac{\partial}{\partial \zeta_1} f(\zeta_1, t) - \frac{D_x}{2} \zeta_1^2 f(\zeta_1, t), \tag{182}$$

$$\frac{\partial}{\partial t} f(v_1, t) \cong [\zeta_1 - (B_z^2 + r_1)v_1] \frac{\partial}{\partial v_1} f(v_1, t) - \frac{D_x}{2} \zeta_1^2 f(v_1, t), \tag{183}$$

$$\frac{\partial}{\partial t} f(\zeta_2, t) \cong -k_2 v_2 \frac{\partial}{\partial \zeta_2} f(\zeta_2, t) - \frac{D_y}{2} \zeta_2^2 f(\zeta_2, t), \tag{184}$$

$$\frac{\partial}{\partial t} f(v_2, t) \cong [\zeta_2 - (B_z^2 + r_2)v_2] \frac{\partial}{\partial v_2} f(v_2, t) - \frac{D_y}{2} \zeta_2^2 f(v_2, t). \tag{185}$$

From the above equations in the steady state, we calculate $f^{st}(\zeta_1, t)$, $f^{st}(v_1, t)$ and $f^{st}(\zeta_2, t)$, $f^{st}(v_2, t)$ as

$$f^{st}(\zeta_1, t) = exp[-\frac{D_x}{2k_1 v_1} \zeta_1 v_1^2], \quad f^{st}(v_1, t) = exp[\frac{D_x}{2[\zeta_1 - B_z^2 v_1]} \frac{v_1^3}{3}], \tag{186}$$

$$f^{st}(\zeta_2, t) = exp[-\frac{D_y}{2k_2 v_2} \zeta_2 v_2^2], \quad f^{st}(v_2, t) = exp[\frac{D_y}{2[\zeta_2 - B_z^2 v_2]} \frac{v_2^3}{3}]. \tag{187}$$

The Fourier transforms of the probability densities $f(\zeta_1, t), f(v_1, t)$ and $f(\zeta_2, t), f(v_2, t)$ are derived as

$$f(\zeta_1, t) = \Theta[t - \zeta_1/k_1 v_1] f^{st}(\zeta_1, t), \quad f(v_1, t) = \Theta[t + v_1/[(\zeta_1 - (B_z^2 + r_1)v_1]] f^{st}(v_1, t), \tag{188}$$

$$f(\zeta_2, t) = \Theta[t - \zeta_2/k_2 v_2] f^{st}(\zeta_2, t), \quad f(v_2, t) = \Theta[t + v_2/[(\zeta_2 - (B_z^2 + r_2)v_2]] f^{st}(v_2, t). \tag{189}$$

We calculate $f(\zeta_1, v_1, t)$ and $f(\zeta_2, v_2, t)$ from Eqs. (188) and (189) as

$$f(\zeta_1, v_1, t) = f(\zeta_1, t) f(v_1, t) = exp[-\frac{D_x t^3}{6} \zeta_1^2 - \frac{D_x t}{2} v_1^2], \tag{190}$$

$$f(\zeta_2, v_2, t) = f(\zeta_2, t) f(v_2, t) = exp[-\frac{D_y t^3}{6} \zeta_2^2 - \frac{D_y t}{2} v_2^2]. \tag{191}$$

By using the inverse Fourier transforms, $f(x,t)$, $f(v_x,t)$ and $f(y,t)$, $f(v_y,t)$ are, respectively, presented by

$$f(x,t) = \left[2\pi \frac{D_x t^3}{3}\right]^{-\frac{1}{2}} exp[-\frac{3x^2}{2D_x t^3}], \quad f(v_x,t) = [2\pi D_x t]^{-1/2} exp[-\frac{v_x^2}{2D_x t}] \tag{192}$$

$$f(y,t) = \left[2\pi \frac{D_y t^3}{3}\right]^{-\frac{1}{2}} exp[-\frac{3y^2}{2D_y t^3}], \quad f(v_y,t) = [4\pi D_y t]^{-1/2} exp[-\frac{v_y^2}{2D_y t}]. \tag{193}$$

The mean squared displacement and the mean squared velocity for $f(x,t)$, $f(v_x,t)$ and $f(y,t)$, $f(v_y,t)$ are, respectively, calculated as

$$< x^2(t) > = \frac{D_x}{3} t^3, \quad < v_x^2(t) > = D_x t, \tag{194}$$

$$< y^2(t) > = \frac{D_y}{3} t^3, \quad < v_y^2(t) > = D_y t. \tag{195}$$

## 5. Statistical quantities and moments

In this section, we calculate the non-Gaussian parameter, the correlation coefficient, the entropy, and the combined entropy for the displacement and the velocity. First of all, the non-Gaussian parameter for displacement

and velocity are, respectively, given by

$$K_x = <x^4>/3<x^2>^2, \quad K_{v_x} = <v_x^4>/3<v_x^2>^2, \tag{196}$$

$$K_y = <y^4>/3<y^2>^2, \quad K_{v_y} = <v_y^4>/3<v_y^2>^2. \tag{197}$$

The non-Gaussian parameter means the degree of tailedness in the probability distribution function of a real-valued variable and provides insight into specific characteristics of a probability distribution in the probability theory and statistics. We introduce the correlation coefficient as

$$\rho_{x,v_x} = x_0 v_{x0}/\sigma_x \sigma_{v_x}, \quad \rho_{y,v_y} = y_0 v_{y0}/\sigma_y \sigma_{v_y} \tag{198}$$

Correlation coefficient $\rho_{x,v_x}$ means the statistical quantity describing the strength and direction of a relationship between two variables $x(t)$ and $v_x(t)$. Here, we assume that a passive particle is initially at $x = x_0$ and at $v_x = v_{x0}$. $\sigma_x$, $\sigma_{v_x}$ denote the root-mean-squared displacement and the root-mean-squared velocity of the joint probability density. The entropies $S(x,t)$ and $S(v_x,t)$ are, respectively, calculated as

$$S(x,t) = -f(x,t)lnf(x,t), \quad S(v_x,t) = -f(v_x,t)lnf(v_x,t), \tag{199}$$

$$S(y,t) = -f(y,t)lnf(y,t), \quad S(v_y,t) = -f(v_y,t)lnf(v_y,t). \tag{200}$$

and the combined entropy is defined by

$$S(x,v_x,t) = -f(x,v_x,t)lnf(x,v_x,t), \quad S(y,v_y,t) = -f(y,v_y,t)lnf(y,v_y,t), \tag{201}$$

The entropy refers to a numerical representation of the reliability or quantity of information that has a probability distribution; in a probability distribution, as the probability of a particular value increases and the probability of the remaining value decreases, the entropy decreases.

In multiplying and integrating both sides of the Fokker–Plank equations by $x^m v^n$, the moment equations for a charged particle from Eq. (9) and Eq. (10) are expressed as

$$\frac{d}{dt}\mu_{m,n} = -m\mu_{m-1,n+1} + B_z v_{y0} n\mu_{m,n-1} + B_z^2 n\mu_{m,n} + D_x n(n-1)\mu_{m-2,n}, \tag{202}$$

$$\frac{d}{dt}\bar{\mu}_{m,n} = -m\bar{\mu}_{m-1,n+1} - B_z v_{x0} n\bar{\mu}_{m,n-1} + B_z^2 n\bar{\mu}_{m,n} + D_x n(n-1)\bar{\mu}_{m-2,n}. \tag{203}$$

Here, $\mu_{m,n} = \int_{-\infty}^{+\infty} dx \int_{-\infty}^{+\infty} dv_x x^m v_x^n p(x,v_x,t)$, and $\bar{\mu}_{m,n} = \int_{-\infty}^{+\infty} dy \int_{-\infty}^{+\infty} dv_y y^m v_y^n p(y,v_y,t)$. The moment equations from Eq. (33) and Eq. (34) for a charged particle with correlated Gaussian forces have

$$\frac{d}{dt}\mu_{m,n} = -m\mu_{m-1,n+1} + B_z v_{y0} n\mu_{m,n-1} + (B_z^2 + r_1)n\mu_{m,n-1} + D_x[-mnb_1(t)\mu_{m-1,n-1} + m(m-1)a_1(t)\mu_{m-2,n}], \tag{204}$$

$$\frac{d}{dt}\bar{\mu}_{m,n} = -m\bar{\mu}_{m-1,n+1} + -B_z v_{x0} n\bar{\mu}_{m,n-1} + (B_z^2 + r_2)n\bar{\mu}_{m,n-1} + D_y[-mnb_2(t)\bar{\mu}_{m-1,n-1} + m(m-1)a_2(t)\bar{\mu}_{m-2,n}]. \tag{205}$$

For the thermal noise, the moment equations from Eq. (86) and Eq. (87) are given by

$$\frac{d}{dt}\mu_{m,n} = -m\mu_{m-1,n+1} + k_1\mu_{m+1,n} + (B_z^2 + r_2)n\mu_{m,n-1} + D_x[mn\frac{t^{2h-1}}{(2h-1)\tau_{th}^{2h-2}}\mu_{m-1,n-1} + m(m-1)\frac{t^{2h}}{2h\tau_{th}^{2h-2}}\mu_{m-2,n}], \tag{206}$$

$$\frac{d}{dt}\bar{\mu}_{m,n} = -m\bar{\mu}_{m-1,n+1} + k_2\bar{\mu}_{m+1,n} + (B_z^2 + r_2)n\bar{\mu}_{m,n-1} + D_y[mn\frac{t^{2h-1}}{(2h-1)\tau_{th}^{2h-2}}\bar{\mu}_{m-1,n-1} + m(m-1)\frac{t^{2h}}{2h\tau_{th}^{2h-2}}\bar{\mu}_{m-2,n}]. \tag{207}$$

For the active noise, the moment equations from Eq. (145) and Eq. (146) are given by

$$\frac{d}{dt}\mu_{m,n} = -m\mu_{m-1,n+1} + k_1\mu_{m+1,n} + (B_z^2 + r_1)n\mu_{m,n-1} + D_x[-mnb_1(t)\mu_{m-1,n-1} + m(m-1)a_1(t)\mu_{m-2,n}], \tag{208}$$

$$\frac{d}{dt}\bar{\mu}_{m,n} = -m\bar{\mu}_{m-1,n+1} + k_2\bar{\mu}_{m+1,n} + (B_z^2 + r_2)n\bar{\mu}_{m,n-1} + D_y[-mnb_2(t)\bar{\mu}_{m-1,n-1} + m(m-1)a_2(t)\bar{\mu}_{m-2,n}]. \tag{209}$$

**Table 1** Values of the non-Gaussian parameter, the correlation coefficient, the entropy, and the combined entropy for the motion of a charged particle with $\vec{B} = B_z \vec{z}$ in the short- and long-time domains. Here, we assume that a charged particle is initially at $x = x_0$, $v_x = v_{x0}$ and at $y = y_0$, $v_y = v_{y0}$.

| Time | $x, v_x$ $y, v_y$ | $K_x, K_{v_x}$ $K_y, K_{v_y}$ | $\rho_{x,v_x}$, $\rho_{y,v_y}$ | $\mu_{m,n}, \bar{\mu}_{m,n}$ | $S(x,t), S(v_x,t)$ $S(y,t), S(v_y,t)$ | $S(x,v_x,t)$ $S(y,v_y,t)$ |
|---|---|---|---|---|---|---|
| short-time | $x$ | $\frac{x_0^4}{D_x^2 B_z^4}t^{-4}+\frac{x_0^2}{D_x B_z^2}t^{-2}$ | $\frac{x_0 v_{x0}}{D_x B_z^4}t^{-2}$ | $\frac{2D_x^2 B_z^6}{3}t^3$ | $\ln D_x B_z^2 t^2$ | $\ln D_x^2 B_z^8 t^4$ |
| | $v_x$ | $\frac{v_{x0}^4}{D_x^2 B_z^{12}}t^{-4}+\frac{v_{x0}^2}{D_x B_z^6}t^{-2}$ | | | $\ln D_x B_z^6 t^2$ | |
| | $y$ | $\frac{y_0^4}{D_y^2 B_z^4}t^{-4}+\frac{y_0^2}{D_y B_z^2}t^{-2}$ | $\frac{y_0 v_{y0}}{D_y B_z^4}t^{-2}$ | $\frac{2D_y^2 B_z^6}{3}t^3$ | $\ln D_y B_z^2 t^2$ | $\ln D_y^2 B_z^8 t^4$ |
| | $v_y$ | $\frac{v_{y0}^4}{D_y^2 B_z^{12}}t^{-4}+\frac{v_{y0}^2}{D_y B_z^6}t^{-2}$ | | | $\ln D_y B_z^6 t^2$ | |
| long-time | $x$ | $\frac{x_0^4}{D_x^2}t^{-2}+\frac{x_0^2}{D_x}t^{-1}$ | $\frac{x_0 v_{x0}}{D_x B_z^3}t^{-3/2}$ | $\frac{2D_x^2 B_z^6}{3}t^3$ | $\ln D_x t$ | $\ln D_x^2 B_z^7 t^3$ |
| | $v_x$ | $\frac{v_{x0}^4}{D_x^2 B_z^{12}}t^{-4}+\frac{v_{x0}^2}{D_x B_z^6}t^{-2}$ | | | $\ln D_x B_z^6 t^2$ | |
| | $y$ | $\frac{y_0^4}{D_y^2}t^{-2}+\frac{y_0^2}{D_y}t^{-1}$ | $\frac{y_0 v_{y0}}{D_y B_z^3}t^{-3/2}$ | $\frac{2D_y^2 B_z^6}{3}t^3$ | $\ln D_y t$ | $\ln D_y^2 B_z^7 t^3$ |
| | $v_y$ | $\frac{v_{y0}^4}{D_y^2 B_z^{12}}t^{-4}+\frac{v_{y0}^2}{D_y B_z^6}t^{-2}$ | | | $\ln D_y B_z^6 t^2$ | |

**Table 2** Values of the non-Gaussian parameter, the correlation coefficient, the entropy, and the combined entropy for the motion of a charged particle with two correlated Gaussian forces within the limits of $t \ll \tau$, $t \gg \tau$ and for $\tau = 0$, where $\tau$ is the correlation time. Here, we assume that a charged particle is initially at $x = x_0, v_x = v_{x0}$ and at $y = y_0, v_y = v_{y0}$.

| Time | $x, v_x$ $y, v_y$ | $K_x, K_{v_x}$ $K_y, K_{v_y}$ | $\rho_{x,v_x}$ $\rho_{y,v_y}$ | $\mu_{m,n}, \bar{\mu}_{m,n}$ | $S(x,t), S(v_x,t)$ $S(y,t), S(v_y,t)$ | $S(x,v_x,t)$ $S(y,v_y,t)$ |
|---|---|---|---|---|---|---|
| $t \ll \tau$ | $x$ | $\frac{x_0^4}{D_x^2}t^{-6}+\frac{x_0^2}{D_x}t^{-3}$ | $\frac{x_0 v_{x0}}{D_x B_z^2}t^{-3}$ | $\frac{2D_x^2 B_z^4}{5\tau_{ac}}t^5$ | $\ln D_x t^3$ | $\ln D_x^2 B_z^4 t^6$ |
| | $v_x$ | $\frac{v_{x0}^4}{D_x^2 B_z^8}t^{-6}+\frac{v_{x0}^2}{D_x B_z^4}t^{-3}$ | | | $\ln D_x B_z^4 t^3$ | |
| | $y$ | $\frac{y_0^4}{D_y^2}t^{-6}+\frac{y_0^2}{D_y}t^{-3}$ | $\frac{y_0 v_{y0}}{D_y B_z^2}t^{-3}$ | $\frac{2D_y^2 B_z^4}{5\tau_{ac}}t^5$ | $\ln D_y t^3$ | $\ln D_y^2 B_z^4 t^6$ |
| | $v_y$ | $\frac{v_{y0}^4}{D_y^2 B_z^8}t^{-6}+\frac{v_{y0}^2}{D_y B_z^4}t^{-3}$ | | | $\ln D_y B_z^4 t^3$ | |
| $t \gg \tau$ | $x$ | $\frac{x_0^4}{D_x^2 B_z^4}t^{-8}+\frac{x_0^2}{D_x B_z^2}t^{-4}$ | $\frac{x_0 v_{x0}}{D_x B_z^3}t^{-7/2}$ | $\frac{2D_x^2 B_z^4}{5\tau_{ac}}t^5$ | $\ln D_x B_z^2 t^4$ | $\ln D_x^2 B_z^6 t^7$ |
| | $v_x$ | $\frac{v_{x0}^4}{D_x^2 B_z^8}t^{-6}+\frac{v_{x0}^2}{D_x B_z^4}t^{-3}$ | | | $\ln D_x B_z^4 t^3$ | |
| | $y$ | $\frac{y_0^4}{D_y^2 B_z^4}t^{-4}+\frac{y_0^2}{D_y B_z^2}t^{-4}$ | $\frac{y_0 v_{y0}}{D_y B_z^3}t^{-7/2}$ | $\frac{2D_y^2 B_z^4}{5\tau_{ac}}t^5$ | $\ln D_y B_z^2 t^4$ | $\ln D_y^2 B_z^6 t^7$ |
| | $v_y$ | $\frac{v_{y0}^4}{D_y^2 B_z^8}t^{-6}+\frac{v_{y0}^2}{D_y B_z^4}t^{-3}$ | | | $\ln D_y B_z^4 t^3$ | |
| $\tau = 0$ | $x$ | $\frac{x_0^4}{D_x^2}t^{-6}+\frac{x_0^2}{D_x}t^{-3}$ | $\frac{x_0 v_{x0}}{D_x B_z^2}t^{-5/2}$ | $\frac{D_x^2 B_z^4}{2\tau_{ac}}t^4$ | $\ln D_x t^3$ | $\ln D_x^2 B_z^4 t^5$ |
| | $v_x$ | $\frac{v_{x0}^4}{D_x^2 B_z^8}t^{-4}+\frac{v_{x0}^2}{D_x B_z^4}t^{-2}$ | | | $\ln D_x B_z^4 t^2$ | |
| | $y$ | $\frac{y_0^4}{D_y^2}t^{-6}+\frac{y_0^2}{D_y}t^{-3}$ | $\frac{y_0 v_{y0}}{D_y B_z^2}t^{-5/2}$ | $\frac{D_y^2 B_z^4}{2\tau_{ac}}t^4$ | $\ln D_y t^3$ | $n D_y^2 B_z^4 t^5$ |
| $v_y$ | | $\frac{v_{y0}^4}{D_y^2 B_z^8}t^{-4} + \frac{v_{y0}^2}{D_y B_z^4}t^{-2}$ | | | | $\ln D_y B_z^4 t^2$ |

**Table 3** Values of the non-Gaussian parameter, the correlation coefficient, the entropy, and the combined entropy for the motion of a charged particle with the trap forces and the thermal noises within the limits of $t \ll \tau$, $t \gg \tau$ and for $\tau = 0$, where $\tau$ is the correlation time. Here, we assume that a charged particle is initially at $x = x_0$, $v_x = v_{x0}$ and at $y = y_0$, $v_y = v_{y0}$.

| Time | $x, v_x$ / $y, v_y$ | $K_x, K_{v_x}$ / $K_y, K_{v_y}$ | $\rho_{x,v_x}$ / $\rho_{y,v_y}$ | $\mu_{m,n}, \bar{\mu}_{m,n}$ | $S(x,t), S(v_x,t)$ / $S(y,t), S(v_y,t)$ | $S(x,v_x,t)$ / $S(y,v_y,t)$ |
|---|---|---|---|---|---|---|
| $t \ll \tau$ | $x$ | $\frac{\tau_{th}^{4h} x_0^4}{D_x^2} t^{-4h-2} + \frac{\tau_{th}^{2h} x_0^2}{D_x} t^{-2h-1}$ | $\frac{\tau_{th}^{2h} x_0 v_{x0}}{D_x k_1} t^{-2h-2}$ | $\frac{h^{-1} D_x^2 k_1^2}{(2h+5)\tau_{th}^{2h-1}} t^{2h+5}$ | $\ln \frac{D_x}{\tau_{th}^{2h}} t^{2h+1}$ | $\ln \frac{D_x^2 k_1^2}{\tau_{th}^{4h}} t^{4h+4}$ |
| | $v_x$ | $\frac{\tau_{th}^{4h} v_{x0}^4}{D_x^2 k_1^4} t^{-4h-6} + \frac{\tau_{th}^{2h} v_{x0}^2}{D_x k_1^2} t^{-2h-3}$ | | | $\ln \frac{D_x k_1^2}{\tau_{th}^{2h}} t^{2h+3}$ | |
| | $y$ | $\frac{\tau_{th}^{4h} y_0^4}{D_y^2} t^{-4h-2} + \frac{\tau_{th}^{2h} y_0^2}{D_y} t^{-2h-1}$ | $\frac{\tau_{th}^{2h} y_0 v_{y0}}{D_y k_2 B_z} t^{-2h-2}$ | $\frac{h^{-1} D_y^2 k_2^2}{(2h+5)\tau_{th}^{2h-1}} t^{2h+5}$ | $\ln \frac{D_y}{\tau_{th}^{2h}} t^{2h+1}$ | $\ln \frac{D_y^2 k_2^2}{\tau_{th}^{4h}} t^{4h+4}$ |
| | $v_y$ | $\frac{\tau_{th}^{4h} v_{y0}^4}{D_y^2 k_2^4} t^{-4h-6} + \frac{\tau_{th}^{2h} v_{y0}^2}{D_y k_2^2} t^{-2h-3}$ | | | $\ln \frac{D_y k_2^2}{\tau_{th}^{2h}} t^{2h+3}$ | |
| $t \gg \tau$ | $x$ | $\frac{\tau_{th}^{4h-2} x_0^4}{D_x^2} t^{-4h-2} + \frac{\tau_{th}^{2h-1} x_0^2}{D_x} t^{-2h-1}$ | $\frac{\tau_{th}^{2h-1/2} x_0 v_{x0}}{D_x k_1 B_z} t^{-2h-2}$ | $\frac{h^{-1} D_x^2 k_1^2}{(2h+5)\tau_{th}^{2h-1}} t^{2h+5}$ | $\ln \frac{D_x}{\tau_{th}^{2h-1}} t^{2h+1}$ | $\ln \frac{D_x^2 k_1^2}{\tau_{th}^{4h-1}} t^{4h+4}$ |
| | $v_x$ | $\frac{\tau_{th}^{4h} v_{x0}^4}{D_x^2 k_1^4} t^{-4h-6} + \frac{\tau_{th}^{2h} v_{x0}^2}{D_x k_1^2} t^{-2h-3}$ | | | $\ln \frac{D_x k_1^2}{\tau_{th}^{2h}} t^{2h+3}$ | |
| | $y$ | $\frac{\tau_{th}^{4h-2} y_0^4}{D_y^2} t^{-4h-2} + \frac{\tau_{th}^{2h-1} y_0^2}{D_y} t^{-2h-1}$ | $\frac{\tau_{th}^{2h-1/2} y_0 v_{y0}}{D_y k_2 B_z} t^{-2h-2}$ | $\frac{h^{-1} D_y^2 k_2^2}{(2h+5)\tau_{th}^{2h-1}} t^{2h+5}$ | $\ln \frac{D_x}{\tau_{th}^{2h-1}} t^{2h+2}$ | $\ln \frac{D_y^2 k_2^2}{\tau_{th}^{4h-1}} t^{4h+4}$ |
| | $v_y$ | $\frac{\tau_{th}^{4h} v_{y0}^4}{D_y^2 k_2^4} t^{-4h-6} + \frac{\tau_{th}^{2h} v_{y0}^2}{D_y k_2^2} t^{-2h-3}$ | | | $\ln \frac{D_y k_2^2}{\tau_{th}^{2h}} t^{2h+3}$ | |
| $\tau = 0$ | $x$ | $\frac{\tau_{th}^{4h} x_0^4}{D_x^2} t^{-4h-2} + \frac{\tau_{th}^{2h} x_0^2}{D_x} t^{-2h-1}$ | $\frac{\tau_{th}^{2h} x_0 v_{x0}}{D_x k_1} t^{-2h-2}$ | $\frac{h^{-1} D_x^2 k_1^2}{(2h+5)\tau_{th}^{2h-1}} t^{2h+5}$ | $\ln \frac{D_x}{\tau_{th}^{2h}} t^{2h+1}$ | $\ln \frac{D_x^2 k_1^2}{\tau_{th}^{4h}} t^{4h+4}$ |
| | $v_x$ | $\frac{\tau_{th}^{2h} v_{x0}^4}{D_x^2 k_1^4} t^{-4h-6} + \frac{\tau_{th}^{2h} v_{x0}^2}{D_x k_1^2} t^{-2h-3}$ | | | $\ln \frac{D_x k_1^2}{\tau_{th}^{2h}} t^{2h+3}$ | |
| | $y$ | $\frac{\tau_{th}^{4h} y_0^4}{D_y^2} t^{-4h-2} + \frac{\tau_{th}^{2h} y_0^2}{D_y} t^{-2h-1}$ | $\frac{\tau_{th}^{2h} y_0 v_{y0}}{D_y k_2} t^{-2h-2}$ | $\frac{h^{-1} D_y^2 k_2^2}{(2h+5)\tau_{th}^{2h-1}} t^{2h+5}$ | $\ln \frac{D_x B_z^2}{\tau_{th}^{2h}} t^{2h+2}$ | $\ln \frac{D_y^2 k_2^2}{\tau_{th}^{4h}} t^{4h+4}$ |
| | $v_y$ | $\frac{\tau_{th}^{2h} v_{y0}^4}{D_y^2 k_2^4} t^{-4h-6} + \frac{\tau_{th}^{2h} v_{y0}^2}{D_y k_2^2} t^{-2h-3}$ | | | $\ln \frac{D_y k_2^2}{\tau_{th}^{2h}} t^{2h+3}$ | |

**Table 4** Values of the non-Gaussian parameter, the correlation coefficient, the entropy, and the combined entropy for the motion of a charged particle with the trap forces and the active noises within the limits of $t \ll \tau$, $t \gg \tau$ and for $\tau = 0$, where $\tau$ is the correlation time. Here, we assume that a charged particle is initially at $x = x_0$, $v_x = v_{x0}$ and at $y = y_0$, $v_y = v_{y0}$.

| Time | $x, v_x$ / $y, v_y$ | $K_x, K_{v_x}$ / $K_y, K_{v_y}$ | $\rho_{x,v_x}$ / $\rho_{y,v_y}$ | $\mu_{m,n}, \bar{\mu}_{m,n}$ | $S(x,t), S(v_x,t)$ / $S(y,t), S(v_y,t)$ | $S(x,v_x,t)$ / $S(y,v_y,t)$ |
|---|---|---|---|---|---|---|
| $t \ll \tau$ | $x$ | $\frac{k_1^2 x_0^4}{D_x^2} t^{-6} + \frac{k_1 x_0^2}{D_x} t^{-3}$ | $\frac{k_1 x_0 v_{x0}}{D_x B_z^2} t^{-3}$ | $\frac{2 D_x^2 B_z^4}{5 k_1 \tau_{ac}} t^5$ | $\ln \frac{D_x}{k_1} t^3$ | $\ln \frac{D_x^2 B_z^4}{k_1^2} t^6$ |
| | $v_x$ | $\frac{k_1^2 v_{x0}^4}{D_x^2 B_z^8} t^{-6} + \frac{k_1 v_{x0}^2}{D_x B_z^4} t^{-3}$ | | | $\ln \frac{D_x B_z^4}{k_1} t^3$ | |
| | $y$ | $\frac{k_2^2 y_0^4}{D_y^2} t^{-6} + \frac{k_2 y_0^2}{D_y} t^{-3}$ | $\frac{k_2 y_0 v_{y0}}{D_y B_z^2} t^{-3}$ | $\frac{2 D_y^2 B_z^4}{5 k_2 \tau_{ac}} t^5$ | $\ln \frac{D_y}{k_2} t^3$ | $\ln \frac{D_y^2 B_z^4}{k_2^2} t^6$ |
| | $v_y$ | $\frac{k_2^2 v_{y0}^4}{D_y^2 B_z^8} t^{-6} + \frac{k_2 v_{y0}^2}{D_y B_z^4} t^{-3}$ | | | $\ln \frac{D_y B_z^4}{k_2} t^3$ | |
| $t \gg \tau$ | $x$ | $\frac{k_1^2 x_0^4}{D_x^2} t^{-6} + \frac{k_1 x_0^2}{D_x} t^{-3}$ | $\frac{k_1 x_0 v_{x0}}{D_x B_z^2} t^{-3}$ | $\frac{2 D_x^2 B_z^4}{5 k_1 \tau_{ac}} t^5$ | $\ln \frac{D_x}{k_1} t^3$ | $\ln \frac{D_x^2 B_z^4}{k_1^2} t^6$ |
|  | $v_x$ | $\frac{k_1^2 v_{x0}^4}{D_x^2 B_z^8} t^{-6} + \frac{k_1 v_{x0}^2}{D_x B_z^4} t^{-3}$ |  |  | $\ln\frac{D_x B_z^4}{k_1} t^3$ |  |
|  | $y$ | $\frac{k_2^2 y_0^4}{D_y^2} t^{-6} + \frac{k_2 y_0^2}{D_y} t^{-3}$ | $\frac{k_2 y_0 v_{y0}}{D_y B_z^2} t^{-3}$ | $\frac{2 D_y^2 B_z^4}{5 k_2 \tau_{ac}} t^5$ | $\ln\frac{D_y}{k_2} t^3$ | $\ln\frac{D_y^2 B_z^4}{k_2^2} t^6$ |
|  | $v_y$ | $\frac{k_2^2 v_{y0}^4}{D_y^2 B_z^8} t^{-6} + \frac{k_2 v_{y0}^2}{D_y B_z^4} t^{-3}$ |  |  | $\ln\frac{D_y B_z^4}{k_2} t^3$ |  |
| $\tau = 0$ | $x$ | $\frac{x_0^4}{D_x^2} t^{-6} + \frac{x_0^2}{D_x} t^{-3}$ | $\frac{x_0 v_{x0}}{D_x} t^{-2}$ | $\frac{2 D_x^2}{3 \tau_{ac}} t^3$ | $\ln D_x t^3$ | $\ln D_x^2 t^4$ |
|  | $v_x$ | $\frac{v_{x0}^4}{D_x^2} t^{-2} + \frac{v_{x0}^2}{D_x} t^{-1}$ |  |  | $\ln D_x t$ |  |
|  | $y$ | $\frac{y_0^4}{D_y^2} t^{-6} + \frac{y_0^2}{D_y} t^{-3}$ | $\frac{y_0 v_{y0}}{D_y B_z^2} t^{-2}$ | $\frac{2 D_y^2}{3 \tau_{ac}} t^3$ | $\ln D_y t^3$ | $\ln D_y^2 t^4$ |
|  | $v_y$ | $\frac{v_{y0}^4}{D_y^2} t^{-2} + \frac{v_{y0}^2}{D_y} t^{-1}$ |  |  | $\ln D_y t$ |  |

## 6. Conclusion

In this work, the double Fourier transform method has been applied to the two-dimensional Vlasov equation describing charged particles subjected to white noise, correlated Gaussian forces, trap forces, and thermal and active noises in a magnetic field. By deriving the associated Fokker–Planck equation, analytical solutions for the joint probability density in different time regimes were obtained. The inclusion of a viscous term suppresses persistent cyclotron oscillations and guarantees the existence of stationary probability densities in the long-time limit. In the long-time regime ($t \gg \tau$), exponentially correlated forces effectively reduce to white noise, and the particle dynamics approaches an Ornstein–Uhlenbeck process in a magnetic field. In this limit, the viscous coefficient explicitly determines both the stationary velocity variance and the effective diffusion coefficient, ensuring convergence of the mean squared displacement.

The main results can be summarized as follows:
(1) In the short-time regime ($t \ll \tau$), where force correlations persist, particle motion resembles ballistic behavior rather than ordinary diffusion, since the random forces are not yet fully decorrelated. In this regime, the mean squared displacement scales as $\sim t^2$, reflecting an approximately constant velocity
(2) In contrast to ordinary Brownian motion, where the mean squared displacement grows linearly with time, the Lorentz force does not significantly suppress diffusion in the short-time limit. However, in the long-time limit, regular diffusive behavior is recovered, with a diffusion coefficient explicitly determined by viscous damping. This demonstrates that the scaling behavior of charged particle motion in a magnetic field depends sensitively on the interplay between magnetic confinement and stochastic or collisional effects.
(3) For charged particles driven by correlated Gaussian noise, the mean squared displacement grows as $\langle x^2(t)\rangle \sim t^4$, indicating a ballistic-like superdiffusive behavior arising from the time-integrated acceleration induced by Lorentz-force–coupled correlated noise.

From the derived Fokker–Planck equation, we obtained approximate analytical expressions for the joint probability density [9–11], which were further validated through numerical simulations based on generalized Langevin equations describing both passive and active particles. These results provide a theoretical framework that can be tested experimentally in future studies of passive and active systems. Extensions of the present model to generalized Langevin equations or equations of motion involving additional types of forces are expected to yield further insights into stochastic transport properties in complex active systems [15-17] and to enable direct comparisons [18-20] with other theoretical, numerical, and experimental investigations.

**Appendix A: Derivation of the time derivative of the joint probability density**

The equations of motion for a charged particle in a magnetic field $\boldsymbol{B} = B_z \hat{z}$ are given by

$$\frac{\partial}{\partial t} x = v_x + \eta_x(t), \quad m\frac{\partial}{\partial t} v_x = -q v_y B_z, \tag{A1}$$

$$\frac{\partial}{\partial t} y = v_y + \eta_y(t), \quad m\frac{\partial}{\partial t} v_y = +q v_x B_z. \tag{A2}$$

Here, the random forces $\eta_x(t)$ and $\eta_y(t)$ depend on the time difference and satisfy $<\eta_i(t)\eta_i(t')> = \frac{D_i}{\tau}\delta(\frac{|t-t'|}{\tau})$ for $i = x, y$. The joint probability density for the displacement $(x, y)$ and the velocity $(v_x, v_y)$ is

defined by $f(x, v_x, y, v_y, t) =< \delta(x - x(t))\delta(v_x - v_x(t)\delta(y - y(t))\delta(v_y - v_y(t)) >$. Taking time derivative of the joint probability density yields

$$\frac{\partial}{\partial t} f = -\frac{\partial}{\partial x} < \frac{\partial x}{\partial t} \delta_x \delta_{v_x} \delta_y \delta_{v_y} > - \frac{\partial}{\partial v_x} < \frac{\partial v_x}{\partial t} \delta_x \delta_{v_x} \delta_y \delta_{v_y} >$$

$$-\frac{\partial}{\partial y} < \frac{\partial y}{\partial t} \delta_x \delta_{v_x} \delta_y \delta_{v_y} > - \frac{\partial}{\partial v_y} < \delta_x \delta_{v_x} \delta_y \delta_{v_y} >, \quad (A3)$$

where $f(x, v_x, y, v_y, t) = f$, $\delta_x = \delta(x - x(t))$, $\delta_{v_x} = \delta(v_x - v_x(t)$, $\delta_y = \delta(y - y(t))$, and $\delta_{v_y} = \delta(v_y - v_y(t))$. Substituting Equations (3) and (4) into Equation (5) and manipulating the method of Ref. [13], we obtain

$$\frac{\partial}{\partial t} f + v_x \frac{\partial}{\partial x} f + v_y \frac{\partial}{\partial y} f + \frac{q}{m} v_y B_z \frac{\partial}{\partial v_x} f - \frac{q}{m} v_x B_z \frac{\partial}{\partial v_y} f = D_x \nabla_x{}^2 f + D_y \nabla_y{}^2 f. \quad (A4)$$

Equation (A4) thus represents the time evolution equation for the joint probability density, corresponding to Eq. (2) in the main text.

**Appendix B: Derivation of the joint probability density**

The joint probability densities $f(x, v_x, t)$ and $f(y, v_y, t)$ for the displacement $(x, y)$ and the velocity $(v_x, v_y)$ are defined by $f(x, v_x, t) =< \delta(x - x(t))\delta(v_x - v_x(t)) >$ and $f(y, v_y, t) =< \delta(y - y(t))\delta(v_y - v_y(t)) >$. Taking time derivatives of the joint probability density yields

$$\frac{\partial}{\partial t} f(x, v_x, t) = -\frac{\partial}{\partial x} < \frac{\partial x}{\partial t} \delta(x - x(t))\delta(v_x - v_x(t)) > - \frac{\partial}{\partial v_x} < \frac{\partial v_x}{\partial t} \delta(x - x(t))\delta(v_x - v_x(t)) >, \quad (B1)$$

$$\frac{\partial}{\partial t} f(y, v_y, t) = -\frac{\partial}{\partial y} < \frac{\partial y}{\partial t} \delta(y - y(t))\delta(v_y - v_y(t)) > - \frac{\partial}{\partial v_y} < \frac{\partial v_y}{\partial t} \delta(y - y(t))\delta(v_y - v_y(t)) >. \quad (B2)$$

Applying the Laplace transform of Eq. (4) and (6), Eq. (4) becomes

$$sv_x(s) - v_{x0} = -B_z v_y(s) - r_1 v_x(s), \quad (B3)$$

and Eq. (6) yields

$$sv_y(s) - v_{y0} = +B_z v_x(s) - r_2 v_y(s). \quad (B4)$$

Here, the initial condition is $x(0) = x_0, v_x(0) = v_{x0}$ and $x(0) = x_0, v_x(0) = v_{x0}$ and $L$ denotes the Laplace transform operator. Solving Eqs. (B3) and Eq. (B4), we obtain

$$v_x(s) = \frac{v_{x0}}{s + r_1} - \frac{B_z}{s + r_1} v_y(s), \quad (B5)$$

$$v_y(s) = \frac{v_{y0}}{s + r_2} + \frac{B_z}{s + r_2} v_x(s). \quad (B6)$$

Taking the inverse Laplace transform, we find

$$v_x(t) = v_{x0} \exp(-r_1 t) - B_z \int dt' \exp[-r_1(t - t')] v_y(t'), \quad (B7)$$

$$v_y(t) = v_{y0} \exp(-r_2 t) + B_z \int dt' \exp[-r_2(t - t')] v_x(t'). \quad (B8)$$

substituting Eqs. (B7) and (B8) into Eqs. (28) and (30) and rearranging Eqs. (31) and (32), we obtain the time evolution equations for $f(x, v_x, t)$ and $f(y, v_y, t)$, given by Eqs. (33) and (34).